\documentclass[a4paper]{article}
\usepackage{graphicx}
\usepackage{dcolumn}
\usepackage{bm}
\usepackage{literat}
\usepackage[left=20mm,top=20mm]{geometry}
\usepackage{amsmath}

\title{Stochastic effects at ripple formation processes in anisotropic systems with multiplicative noise}
\author{D.O.~Kharchenko\footnote{dikh@ipfcentr.sumy.ua}, V.O.~Kharchenko, I.O.~Lysenko,
S.V.~Kokhan\\ \ \\
 \textit{Institute of Applied Physics, Nat.Acad. of Sci. of Ukraine,}\\ \textit{58
		  Petropavlovskaya St., 40030, Sumy, Ukraine}}
\begin{document}
\maketitle
\section*{Abstract}
We study pattern formation processes in anisotropic system governed by the
Kuramoto-Sivashinsky equation with multiplicative noise as a generalization of
the Bradley-Harper model for ripple formation induced by ion bombardment. For
both linear and nonlinear systems we study noise induced effects at ripple
formation and discuss scaling behavior of the surface growth and roughness
characteristics. It was found that the secondary parameters of the ion beam
(beam profile and variations of an incidence angle) can crucially change the
topology of patterns and the corresponding dynamics.

 \textbf{PACS} 05.40.-a, 05.70.Ln, 64.60.l, 68.35.Ct, 79.20.Rf

\section{Introduction} \label{sec:intro}
A fabrication of nanoscale surface structures have attracted a considerable
attention due to their applications in electronics \cite{Rev98}. In the last
five decades many studies have been devoted to understanding the mechanism of
pattern formation and its control during ion-beam sputtering (see, for example
Refs.\cite{NSCA64,BH88,CB95,MB97,DZLW99,AV04,KK04,KYH2009}). Among theoretical
investigations there are a lot of experimental data manifesting a large class
of patterns appeared as result of self-organization process on the surface of a
solid. It was shown experimentally that main properties of pattern formation
and structure of patterns depend on the energetic ion-beam parameters such as
ion flux, energy of deposition, angle of incidence and temperature. Formation
of ripples was investigated on different substrates, i.e. on metals ($Ag$ and
$Cu$) \cite{RBV97,RCBV98} on semiconductors ($Ge$ \cite{CMK94} and $Si$
\cite{EAC99,China2006,MAA07}) on $Sn$ \cite{QZF05}, $InP$ \cite{PMS08}, on
$Cd_2Nb_2O_7$ pyrochlore \cite{LWW06} and other. Height modulations on the
surface induced by ion-beam sputtering result in formation of ripples having
the typical size of 0.1 to 1 $\mu m$ and nanoscale patterns with the linear
size of 35 to 250 $A$ \cite{Sicence99}.

It is well known that orientation of ripples depends on the incidence angle. At
the incidence angles around $\pi/2$ the wave-vector of the modulations is
parallel to the component of the ion beam in the surface plane, whereas at
small incidence angles (close to grazing) the wave-vector is perpendicular to
this component. The orientation of ripples can be controlled by a penetration
depth which is proportional to the deposited energy. Analytical investigations
provided by Cuerno and Barabasi show a possible control of pattern formation
governed by both the incidence angle and penetration depth \cite{CB95,MB97}.
The main theoretical models describing ripple formation are based on results of
the famous works of Bradley and Harper \cite{BH88}, Kardar, Parisi, and Zhang
\cite{KPZ86}, Wolf and Villian \cite{WV90}, Kuramoto and Sivashinsky
\cite{KSeq}. The main mechanisms for pattern formation were set to predict
orientation change of the ripples, formation of holes and dots. These models
were generalized taking into account additive fluctuations leading to
statistical description of the corresponding processes.

Moreover, it was shown that under well defined processing conditions the
secondary ion-beam parameters (beam profile) may lead to different patterns
\cite{ZFT08}. Theoretical predictions including statistical properties of the
beam profile were performed in Ref.\cite{KYH2009}. It was shown that
fluctuations in incident angles result in stochastic description of the ripple
formation with multiplicative noise. Unfortunately, detailed description of
pattern formation in such complicated stochastic systems was not discussed.
Moreover, the problem of understanding the scaling behavior of the surface
characteristics is still opened.

In this article we aim to study ripple (or generally pattern) formation
processes in anisotropic system governed by the corresponding
Kuramoto-Sivashinsky equation which takes into account multiplicative noise
caused by fluctuation of the incidence angle. We consider the linear and
nonlinear models separately and discuss the corresponding phase diagrams in the
space of main beam parameters reduced to the penetration depth and the
incidence angle. Moreover, we present results of the scaling behavior study of
the correlation functions and discuss time dependencies of the roughness and
growth exponents during the system evolution as well as fractal properties of
the surface. It will be shown that multiplicative fluctuations in ripple
formation processes can accelerate/delay surface modulations. We shall show
that both phase diagrams and the scaling exponents crucially depend on the
statistical properties of the beam.

The work is organized as follows. In Section \ref{model} we present the
stochastic model with multiplicative noise. Section \ref{stabAn} is devoted to
the stability analysis of the linear system, where the main phase diagrams are
discussed. The nonlinear stochastic model is studied in Section \ref{NLIN}.
Here we consider the behavior of the main statistical characteristics of the
surface such as distribution of the height field, scaling properties of the
correlation functions. We summarize in Section \ref{sum}.

\section{Model}\label{model}
Let us consider a $d$-dimensional substrate and denote with $\mathbf{r}$ the
$d$-dimensional vector locating a point on it. The surface is described at each
time $t$ by the height $z=h(\mathbf{r},t)$. If we assume that the surface
morphology is changed while ion sputtering, then we can use the model for the
surface growth proposed by Bradley and Harper \cite{BH88} and further developed
by Cuerno and Barabasi \cite{CB95}. We consider the system where the direction
of the ion beam lies in $x-z$ plane at an angle $\theta$ from the normal of the
uneroded surface. Following the standard approach one assumes that an averaged
energy deposited at the surface (let say point $O$) due to the ion arriving at
the point $P$ in the solid follows the Gaussian distribution \cite{BH88}
$E(\mathbf{r})=(\epsilon/(2\pi)^{3/2}\sigma\mu^2)\exp(-z^2/2\sigma^2-(x^2+y^2)/2\mu^2)$;
$\epsilon$ denotes the kinetic energy of the arriving ion, $\sigma$ and $\mu$
are the widths of the distribution in directions parallel and perpendicular to
the incoming beam. Parameters $\sigma$ and $\mu$ depend on the target material
and can vary with physical properties of the target and incident energy. We
consider the simplest case when $\sigma=\mu$. The erosion velocity at the
surface point $O$ is described by the formula $v=p\int_\mathcal{R}{\rm
d}\mathbf{r}\Phi(\mathbf{r})E(\mathbf{r})$, where integration is provided over
the range of the energy distribution of all ions; here $\Phi(\mathbf{r})$ and
$p$ are corrections for the local slope dependence of the uniform flux $J$ and
proportionality constant, respectively \cite{Sigmund1973}. The general
expression for the local flux for surfaces with non-zero local curvature is
\cite{MB04}: $\Phi(x,y,h)=J\cos\left(\arctan\left[\sqrt{(\nabla_x
h)^2+(\nabla_y h)^2}\right]\right)$.
 Hence, the dynamics of the surface height is defined by the relation
$\partial_t h\simeq-v(\theta-\nabla_x h, \nabla_x^2 h, \nabla_y^2 h)$ and is
given by the equation $\partial_t h\simeq-v(\theta)\sqrt{1+(\nabla h)^2}$,
where $0<\theta<\pi/2$ \cite{BH88,CT94,CB95,KPZ86,MB97}. The linear term
expansion gives $\partial_t
h=-v_0+\gamma\nabla_xh+\nu_\alpha\nabla_{\alpha\alpha}^2h$; where
$\nabla=\partial/\partial \mathbf{r}$, $\nabla_\alpha=\partial/\partial
\alpha$, $\alpha=\{x,y\}$. Here $v_0$ is the surface erosion velocity;
$\gamma=\gamma(\theta)$ is a constant that describes the slope depending
erosion; $\nu_\alpha=\nu_\alpha(\theta)$ is effective surface tension generated
by erosion process in $\alpha$ direction.

If one assumes that the surface current is driven by differences in chemical
potential $\mu$, then the evolution equation for the field $h$ should take into
account the term $-\nabla\cdot \mathbf{j}_s$ in the right hand side, where
$\mathbf{j}_s= K\nabla(\nabla^2h)$ is the surface current; $K>0$ is the
temperature dependent surface diffusion constant. If the surface diffusion is
thermally activated, then we have $K=D_s\kappa\rho/n^2T$, where
$D_s=D_0e^{-E_a/T}$ is the surface self-diffusivity ($E_a$ is the activation
energy for surface diffusion), $\kappa$ is the surface free energy, $\rho$ is
the areal density of diffusing atoms, $n$ is the number of atoms per unit
volume in the amorphous solid. This term in the dynamical equation for $h$ is
relevant in high temperature limit which will be studied below.

Quantities $v_0$, $\gamma$, $\nu_\alpha$ are functions of the angle $\theta$
only, not the temperature. Assuming that the surface varies smoothly, next we
neglect spatial derivatives of the height $h$ of third and higher orders in the
slope expansion. Taking into account nonlinear terms in the slope expansion of
the surface height dynamics, we arrive at the equation for the quantity
$h'=h+v_0t$ of the form \cite{BH88,CB95}
\begin{equation}\label{eq1}
\partial_t h=\gamma\nabla_xh+\nu_\alpha\nabla_{\alpha\alpha}^2h+\frac{\lambda_\alpha}{2}(\nabla_\alpha h)^2-
K\nabla^2(\nabla^2h),
\end{equation}
where we drop the prime for convenience. Coefficients in Eq.(\ref{eq1}) are
defined in Ref.\cite{CB95} and read
\begin{eqnarray}
 &s=\sin\theta,\quad c=\cos\theta,\quad  a_\sigma=a/\sigma, \quad  F\equiv (\epsilon p
J/\sqrt{2\pi})\exp(-a_\sigma^2/2),\nonumber\\
 &\gamma=\frac{F}{\sigma} s(a_\sigma^2c^2-1),\nonumber\\
&\lambda_x=\frac{F}{\sigma}c(a_\sigma^2(3s^2-c^2)-a_\sigma^4s^2c^2),\quad
	 \lambda_y=-\frac{F}{\sigma}c(a_\sigma^2c^2),\nonumber\\
 &\nu_x=\frac{F}{2}a_\sigma(2s^2-c^2-a_\sigma^2s^2c^2),\quad
\nu_y=-\frac{F}{2}a_\sigma c^2\nonumber
\end{eqnarray}
Here all control parameters are defined through the ion penetrate distance $a$,
the incidence angle $\theta$, the flux $J$ and the kinetic energy $\epsilon$.
It is known \cite{MB04} that the penetration depth depends on the target
material properties and the incoming ion energy $\epsilon$: $a\approx
\epsilon^{2m}/n C_m$, where $n$ is the target atom density, $C_m$ is the
constant depending on the interatomic interaction potential \cite{Sigmund69},
$m\approx 1/2$ for intermediate energies (from 1 to 100 keV). Equation
(\ref{eq1})is known as the noiseless anisotropic Kuramoto-Sivashinsky equation
\cite{KSeq}

It was shown \cite{BH88} that the linearized dynamical equation (\ref{eq1})
admits a solution of the form $h(x,y,t)=A\exp(i(k_x x+k_y y-\omega t)-r t)$,
where $\omega=-\gamma(\theta) k_x$ is the frequency, $r=-(\nu_x(\theta)
k_x^2+\nu_y(\theta) k_y^2)-K(k_x^2+k_y^2)^2$ is the parameter responsible for a
stability of the solution. During the system evolution a selection of
wave-numbers responsible for ripple orientation occurs. The selected
wave-number is $k^2_\alpha=|\nu_\alpha|/2K$, where $\alpha$ refers to the
direction ($x$ or $y$) along which the associated $\nu_\alpha$ has smaller
value.

For the noiseless nonlinear model (\ref{eq1}) it was shown that due to the sets
$\nu_\alpha$ and $\lambda_\alpha$ are the functions of the incidence angle
$\theta\in [0,\pi/2]$ there are three domains in the phase diagram
$(a_\sigma,\theta)$ where $\nu_x$ and $\lambda_x$ changes their signs,
separately \cite{CB95}. It results in ripples formation in different direction
$x$ or $y$ varying $a_\sigma$ or $\theta$.

To describe an evolution of the surface in more realistic conditions one should
take into account that the bombarding ions reach the surface stochastically,
i.e. at random position and time; generally, it can reach the surface at random
angle lying in the vicinity of the incidence angle $\theta$. Most of models
proposed to describe ripple formation due to the ion sputtering process
incorporates additive fluctuations $\xi(\mathbf{r},t)$ that takes into account
stochastic nature of arriving ions (see for example
Refs.\cite{CB95,DZLW99,China2006}). From the mathematical viewpoint such
stochastic source results in spreading the patterns and makes possible
statistical description of the system. If this term is assumed as a Gausisan
white noise in time and space it can not change the system behavior crucially
\cite{Gardiner,Garcia}.

If one supposes that the ion beam is composed of ions distributed with
different incidence angles, then we have three possible cases \cite{KYH2009}:
(i) homogeneous beam when the erosion velocity depends upon random ion beam
parameters and the average velocity is defined through the distribution
function over beam directions; (ii) temporally fluctuating homogenous beam when
the direction of illumination constitutes a stationary, temporally homogeneous
stochastic process; (iii) spatio-temporally fluctuating beam when the
directions of ions form a homogeneous and stationary field. In
Ref.\cite{KYH2009} authors consider the case (iii) under assumption of the
Gaussian distribution of a beam profile centered at a fixed angle $\theta_0$.
Such model means that the fluctuation term that can appear in the dynamical
equation for the field $h$ is some kind of a multiplicative noise (with
intensity depending on the field $h$). Unfortunately only general perspectives
were reported for the nonlinear model, whilst main results relate to studying
the linear model behavior. From the naive consideration one can expect that the
multiplicative noise can qualitatively influence on the dynamics of ripple
formation in the nonlinear system.

In present article we aim to consider the general problem of the ripple
formation under assumption of Gaussian distribution of the beam profile around
$\theta_0$ in the framework of the model given by Eq.(\ref{eq1}) following the
approach proposed in Ref.\cite{KYH2009}. To describe the model we start from
Eq.(\ref{eq1}) which can be rewritten in the form $\partial_t h=f(\theta,
\nabla_\alpha h)$, where $f$ is a deterministic force. Considering small
deviations from the fixed angle $\theta_0$ we can expand the function
$f(\theta, \nabla_\alpha h)$ in the vicinity of $\theta_0$. Therefore, for the
right hand side we get $f=f_0+(\partial f/\partial
\theta)|_{\theta=\theta_0}\delta \theta$ and assume that $\delta\theta$ is a
stochastic field, i.e. $\delta\theta=\delta\theta(\mathbf{r},t)$. Assuming
Gaussian properties for the stochastic component $\delta\theta$, we set
\begin{equation}
 \langle \delta\theta (\mathbf{r},t)\rangle=0, \quad
 \langle \delta\theta (\mathbf{r},t)\delta\theta ( \mathbf{r}',t')\rangle= 2D \Sigma C_r(
 \mathbf{r}-\mathbf{r}')C_t(t-t'),
\end{equation}
where $D$ is the parameter depending on the beam characteristics such as $J$,
$\epsilon$, $p$, $a$, $\sigma$; $\Sigma$ is the noise intensity characterizing
 dispersion of $\delta\theta$; $C_r$ and $C_t$ are spatial and temporal
correlation functions of the noise $\delta\theta$. In further consideration we
assume that $\delta\theta$ is the quasi-white noise in time with $C_t(t-t')\to
\delta(t-t')$ and colored in space, i.e.
$C_r(\mathbf{r}-\mathbf{r}')=(\sqrt{2\pi
r_c^2})^{-d}\exp(-(\mathbf{r}-\mathbf{r}')^2/2r_c^2)$, where $r_c$ is the
correlation radius of fluctuations. At $\Sigma=0$ no fluctuations in the beam
directions (incidence angle) are realized (pure deterministic case).

Therefore, expanding coefficients at spatial derivatives in Eq.(\ref{eq1}) we
arrive at the Langevin equation of the form
\begin{equation}\label{eq2}
\partial_t h=\gamma_0\nabla_xh+\nu_{\alpha 0}\nabla_{\alpha\alpha}^2h+\frac{\lambda_{\alpha 0}}{2}(\nabla_\alpha h)^2-
K\nabla^2(\nabla^2h) +\left[\gamma_1\nabla_xh+\nu_{\alpha
1}\nabla_{\alpha\alpha}^2h+\frac{\lambda_{\alpha 1}}{2}(\nabla_\alpha
h)^2\right]\delta\theta,
\end{equation}
where $\gamma_0=\gamma(\theta_0)$, $\nu_{\alpha 0}=\nu_{\alpha}(\theta_0)$,
$\lambda_{\alpha 0}=\lambda_{\alpha}(\theta_0)$,
$\gamma_1=\partial_\theta\gamma|_{\theta=\theta_0}$, $\nu_{\alpha
1}=\partial_\theta\nu_{\alpha}|_{\theta=\theta_0}$, $\lambda_{\alpha
1}=\partial_\theta\lambda_{\alpha}|_{\theta=\theta_0}$. The parameter $D$ is
reduced to the constant $F$, that means that multiplicative fluctuations
appears only if the system is subjected to ion beam with $F\ne0$. Therefore,
the stochastic system is described by the anisotropic Kuramoto-Sivashinsky
equation with the multiplicative noise.

\section{Stability analysis of the linear model}\label{stabAn}

It is known that transitions between two macroscopic phases in a given system
occur due to the loss of stability of the state for the certain values of the
control parameters. In the case of stochastic systems the liner stability
analysis needs to be performed on a statistical moment of the perturbed state.
We will now perform the stability analysis for the system with multiplicative
fluctuations. To that end we average the Langevin equation (\ref{eq2}) over
noise and obtain
\begin{equation}\label{eq2}
\begin{split}
\partial_t \langle h\rangle =&\gamma_0\nabla_x\langle h\rangle+\nu_{\alpha 0}\nabla_{\alpha\alpha}^2\langle h\rangle+
\frac{\lambda_{\alpha 0}}{2}\langle(\nabla_\alpha h)^2\rangle-
K\nabla^2(\nabla^2\langle h\rangle)\\
&+\left<\left[\gamma_1\nabla_xh+\nu_{\alpha
1}\nabla_{\alpha\alpha}^2h+\frac{\lambda_{\alpha 1}}{2}(\nabla_\alpha
h)^2\right]\delta\theta\right>.
\end{split}
\end{equation}
The last term can be calculated using the Novikov theorem \cite{Novikov}. From
a formal representation one has
\begin{equation}
\langle \mathcal{R} \delta\theta(x,y;t)\rangle=\int{\rm d}t'\int{\rm
d}x'\int{\rm d}y'\langle \delta\theta(x,y;t)\delta\theta(x',y';t')\rangle\left<
\frac{\delta\mathcal{R}}{\delta(\delta\theta(x'y';t'))}\right>,
\end{equation}
where $\mathcal{R}$ is the functional, $\delta/\delta(\delta\theta)$ is the
variational derivative. The integration is carried out over the whole range of
$x'$, $y'$ and $t'$. For our model one has
$\mathcal{R}=\gamma_1\nabla_xh+\nu_{\alpha
1}\nabla_{\alpha\alpha}^2h+\frac{\lambda_{\alpha 1}}{2}(\nabla_\alpha h)^2$.
The variational derivative can be computed with the help of the relation
$\frac{\delta\mathcal{R}}{\delta(\delta\theta)}=\frac{\partial
\mathcal{R}}{\partial h}\left(\frac{\partial h}{\partial \delta\theta
}\right)_{\alpha=\alpha'}$, where the second term is obtained from the formal
solution of the Langevin equation (\ref{eq2}). It follows that the response
function takes the form
\begin{equation}
\left(\frac{\partial h}{\partial \delta\theta
}\right)_{\alpha=\alpha'}=\gamma_1\nabla_xh\delta(x-x')+\delta(\alpha-\alpha')\left\{\nu_{\alpha
1}\nabla_{\alpha\alpha}^2h+\frac{\lambda_{\alpha 1}}{2}(\nabla_\alpha
h)^2\right\}.
\end{equation}

As a result the variational derivative can be written as follows
\begin{equation}
\begin{split}
\frac{\delta\mathcal{R}}{\delta (\delta\theta)}=
 &\gamma_1\nabla_x\left[\gamma_1\nabla_x h\delta(x-x')+\delta(\alpha-\alpha')\left\{
	   \nu_{\alpha 1}\nabla_{\alpha\alpha}^2h +\frac{\lambda_{\alpha 1}}{2}(\nabla_\alpha h)^2 \right\}\right]\\
 +&\nu_{\alpha 1}\nabla_{\alpha\alpha}^2 \left[\gamma_1\nabla_x h\delta(x-x')+\delta(\beta-\beta')\left\{
	   \nu_{\beta1}\nabla_{\beta\beta'}^2h +\frac{\lambda_{\beta 1}}{2}(\nabla_\beta h)^2
	   \right\}\right]\\
 +&\lambda_{\alpha 1}(\nabla_\alpha h)\nabla_\alpha\left[\gamma_1\nabla_x h\delta(x-x')+\delta(\beta-\beta')\left\{
	   \nu_{\beta1}\nabla_{\beta\beta'}^2h +\frac{\lambda_{\beta 1}}{2}(\nabla_\beta h)^2
	   \right\}\right].
\end{split}
\end{equation}

Let us consider the stability of the linear system. From the relation obtained
it follows that terms with coefficients $\lambda_{\alpha1}$ lead to the
nonlinear contribution, and hence can be neglected at this stage. Therefore,
reduced expression is of the form
\begin{equation}
\begin{split}
\frac{\delta\mathcal{R}}{\delta (\delta\theta)}=
  &\gamma_1^2\nabla_x\left[\nabla_x h\delta(x-x')\right]+\gamma_1\nu_{\alpha1}\left\{\nabla_x\left[\nabla_{\alpha\alpha}^2h\delta(\alpha-\alpha')\right]+ \nabla^2_{\alpha\alpha}[\nabla_x h\delta(x-x')]\right\}\\
  +& \nu_{\alpha1}\nu_{\beta 1}\nabla^2_{\alpha\alpha}\left[\nabla^2_{\beta\beta}h\delta(\alpha-\beta')\right].
\end{split}
\end{equation}

To perform next calculations we assume that the spatial correlation function
for fluctuations can be decomposed as $C_r(\mathbf{r})=C_x(x)C_y(y)$ with
maximum at $\alpha=\alpha'$, where $C(0)\equiv C_x(0)=C_y(0)$ and
$C''|_{\alpha=\alpha'}\equiv\partial^2_{xx} C_x|_{x=x'}=\partial^2_{yy}
C_y|_{y=y'}$, with $C''|_{\alpha=\alpha'}<0$. Then, integrating over $t'$, and
$x'$ and $y'$ (by parts), we obtain the expression for the decomposed
correlator:
\begin{equation}
\begin{split}
\left<\left[\gamma_1\nabla_xh+\nu_{\alpha
1}\nabla_{\alpha\alpha}^2h\right.\right.&\left.\left.+(\lambda_{\alpha
1}/2)(\nabla_\alpha h)^2]\right]\delta\theta\right>\simeq\\
 &\left\{\nu_{\alpha1}^2
C''|_{\alpha=\alpha'}\nabla_{\alpha\alpha}^2 +
\gamma_1^2C(0)\nabla_{xx}^2+C(0)(\nu_{\alpha1}\nabla_{\alpha\alpha}^2)^2\right.\\
&\left.+\gamma_1\nu_{1x}\left[
C''|_{\alpha=\alpha'}\nabla_x+C(0)\nabla_{xxx}^3\right]+\gamma_1\nu_{\alpha1}C(0)\nabla_{\alpha\alpha}^2\nabla_x
\right\}\langle h\rangle.
\end{split}
\end{equation}

Finally, we can rewrite the linearized evolution equation for the average
$\langle h\rangle$ in the standard form:
\begin{equation}\label{avh}
\partial_t \langle h\rangle =\widehat{\gamma_{ef}}\langle h\rangle+\widehat{\nu_{ef}}\langle
h\rangle-\widehat{K_{ef}}\langle h\rangle,
\end{equation}
where the following notations are used
\begin{equation}
\begin{split}
 &\widehat{\gamma_{ef}}\equiv(\gamma_{0}+\gamma_{1}\Sigma\left[\nu_{x1}C''|_{\alpha=\alpha'}+\nu_{x1}C(0)\nabla_{xx}^2+\nu_{\alpha1}C(0)\nabla_{\alpha\alpha}^2\right] )\nabla_x,\\
 &\widehat{\nu_{ef}}\equiv(\nu_{\alpha0}+\Sigma C''|_{\alpha=\alpha'}\nu^2_{\alpha1})\nabla_{\alpha\alpha}^2+\gamma_1^2\Sigma C(0)\nabla_{xx}^2,\\
 &\widehat{K_{ef}}\equiv
 -K(\nabla_{\alpha\alpha}^2)^2+\Sigma C(0)(\nu_{\alpha1}\nabla_{\alpha\alpha}^2)^2.
\end{split}
\end{equation}
It is easy to see that Eq.(\ref{avh}) admits a solution of the form $\langle
h\rangle=A\exp(i(k_xx+k_yy-\omega t)+rt)$. Indeed, substituting it into
Eq.(\ref{avh}) and separating real and imaginary parts we found
\begin{equation}
\begin{split}
 &\omega=-k_x(\gamma_0+\gamma_{1}\nu_{x1}\Sigma  C''|_{x=x'})+k_x\gamma_{1}\nu_{x1}\Sigma  C(0)(k_x^2+k_y^2),\\
 &r=-k_x^2\Gamma_x-k_y^2\Gamma_y-K(k_x^2+k_y^2)^2+\Sigma  C(0)(\nu_{x1}^2k_x^2+\nu_{y1}^2k_y^2)^2,\\
 &\Gamma_x\equiv \nu_{x0}+\nu_{x1}^2\Sigma  C''|_{x=x'}+\gamma_{1}^2\Sigma  C(0),\quad \Gamma_y\equiv \nu_{y0}+\nu_{y1}^2\Sigma  C''|_{y=y'}.
\end{split}
\end{equation}
It follows that if $\Gamma_\alpha<0$, then there will be a range of low
frequencies that will grow exponentially. From our model one can see that as
$\nu_{y0}<0$ and $C''<0$ with $C(0)>0$ the quantity $\Gamma_y$ is always
negative. Therefore, instability along $y$ direction will always appear. The
quantity $\Gamma_x$ can change sign as control parameters $\theta$ and
$a_\sigma$ and noise characteristics vary. It means that instability in $x$
direction can appear at same incidence angles and penetration depths. Moreover,
the statistical characteristics of the noise reduced to the spatial correlation
length $r_c$ and the intensity $\Sigma$ governing the total stability of the
solution can change the system behavior drastically.

\begin{figure}
\centering
 a)\includegraphics[width=70mm]{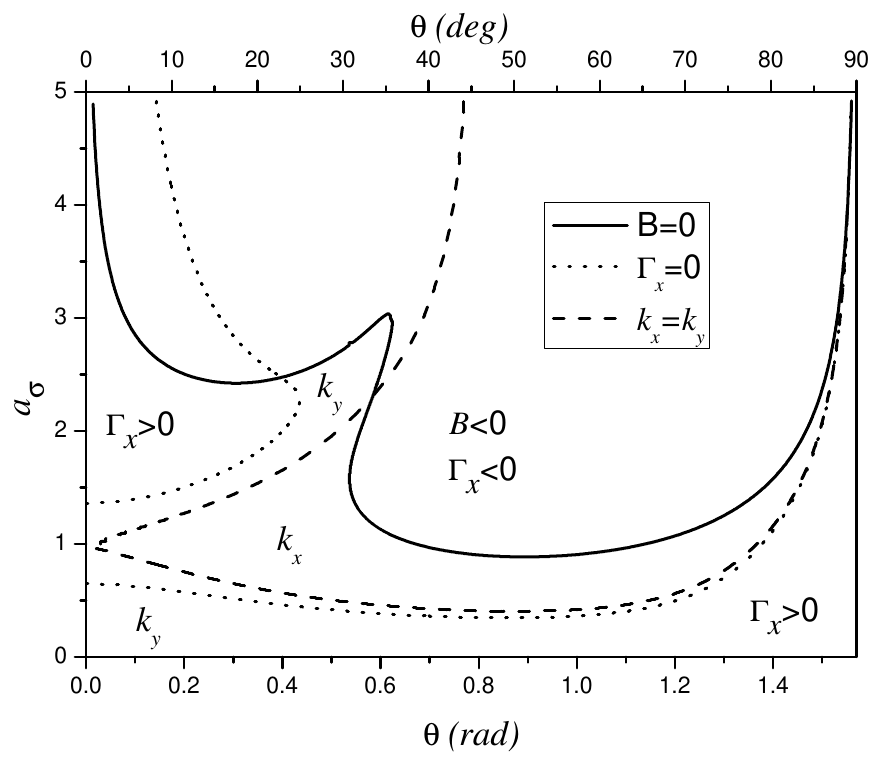} \ b)\includegraphics[width=87mm]{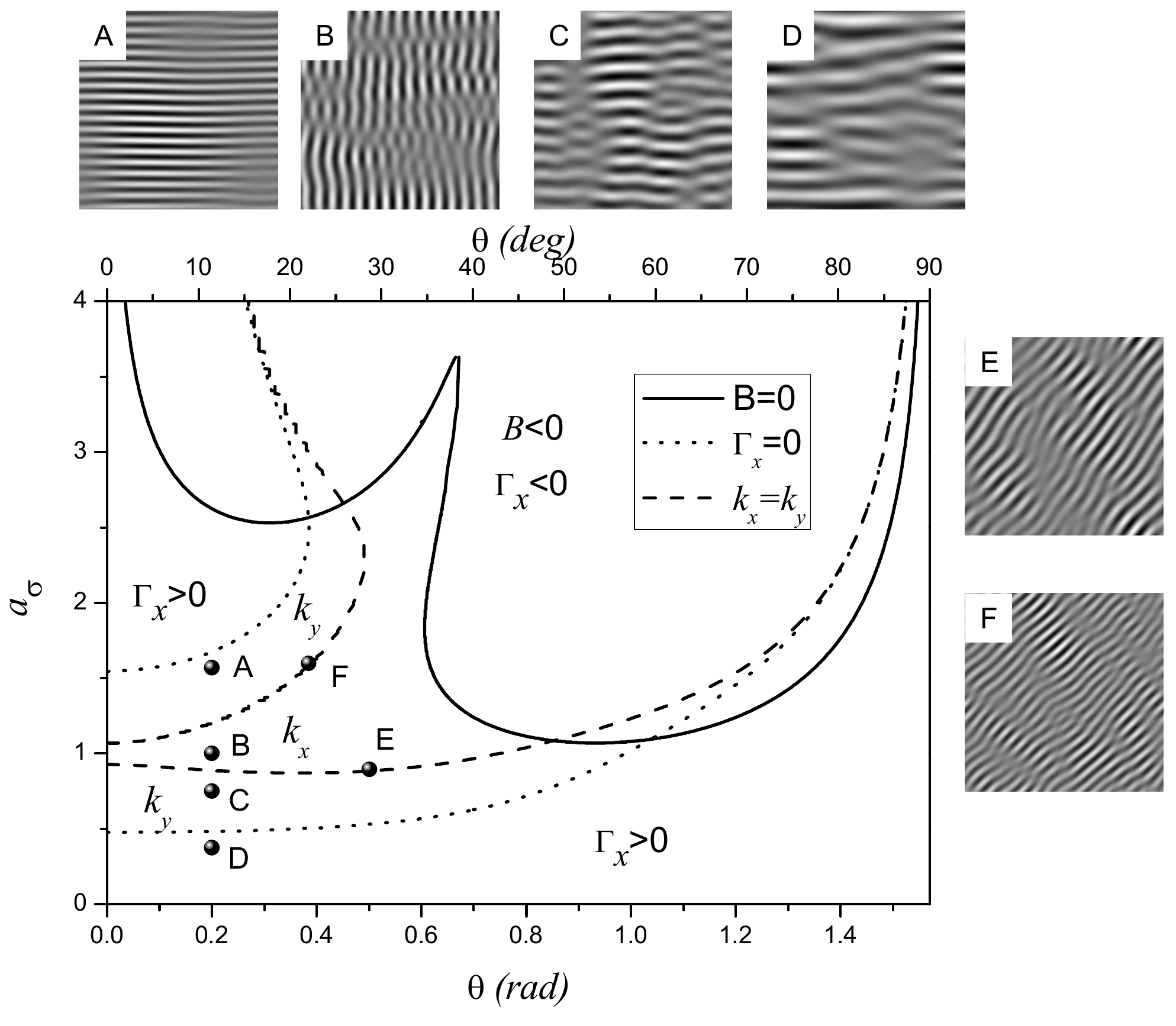}
\caption{ Phase diagrams for the pattern selection at in the system with
multiplicative noise with $\Sigma=1$ (in domains denoted with $k_\alpha$ where
 $\alpha\in\{x,y\}$ patterns with wave-vector
$\mathbf{k}=|k_\alpha|\hat\alpha$ are selected; plots (a), (b) correspond to
$r_c=0.65$, $r_c=1$, respectively).\label{fig1}}
\end{figure}

Stability change of the anisotropic system with an additive noise was discussed
earlier \cite{CB95}. Let us consider stability change in the
 system with  the multiplicative noise. In Figures \ref{fig1}a,b we plot the
corresponding phase diagrams at fixed noise intensity $\Sigma$ and different
correlation radius $r_c$. Here dotted lines limit domains of the stability of
the system at low frequencies and relate to the case $\Gamma_x=0$. Solid line
divides the space of $a_\sigma$ and $\theta$ where parameter $B\equiv
2K-4\Sigma C(0)\nu_{\alpha1}^4$ takes zero values at $k_x=k_y$. This parameter
is responsible for the stability of the system at large wave-numbers. It is
known that observable/selected ripples correspond to wave-numbers with
$k^2_\alpha=|\Gamma|/B$ where $B>0$ and $\Gamma=\min[\Gamma_x, \Gamma_y]$.
Dashed lines in Fig.\ref{fig1} correspond to the system parameters where
$k_x=k_y$. In domains denoted with the corresponding wave-number $k_x$ or $k_y$
ripples have the orientation in $x$ or in $y$ direction, respectively. As it
follows from our linear stability analysis, orientation of ripples can be
controlled varying both the penetration depth $a_\sigma$ and the angle of
incidence $\theta$ at fixed $\Sigma$ and $r_c$. Comparing plots in
Fig.\ref{fig1}a and in Fig.\ref{fig1}b one can see that the statistical
properties of the noise $\delta\theta$ are responsible for the change of the
system behavior. Indeed, at small correlation radius of the angle fluctuations
the domain of the system instability at fixed $a_\sigma=1$ is bigger than at
large $r_c$. Moreover, the variation of quantity $r_c$ can lead to a decrease
of the system parameters where ripples oriented along $k_x$ are observed. It is
interesting to note that at large $r_c$ at fixed interval of the incidence
angles $\theta$ a reorientation of ripples can be found varying parameter
$a_\sigma$ related to the deposited energy of the beam. Indeed, in the interval
of $\theta$ lying between the abscissa of point $E$ and abscissa of point $F$
some kind of reentrance is observable: at small $a_\sigma$ (below the bottom
dashed line where $k_x=k_y$) the ripples are oriented along $k_y$; in the
intermediate domain of $a_\sigma$ (between two dashed lines) the ripples are
oriented along $k_x$; at large $a_\sigma$ the ripples are oriented along $k_y$
again (see snapshots for points $A-D$). The same situation is realized at fixed
$a_\sigma$ when the incidence angle varies. For the system parameters related
to the dashed lines (see points $E$, $F$) the ripples are characterized by
$k_y=k_y$ with the orientation angle $\pi/4$.

\begin{figure}
\centering
 a) \includegraphics[width=80mm]{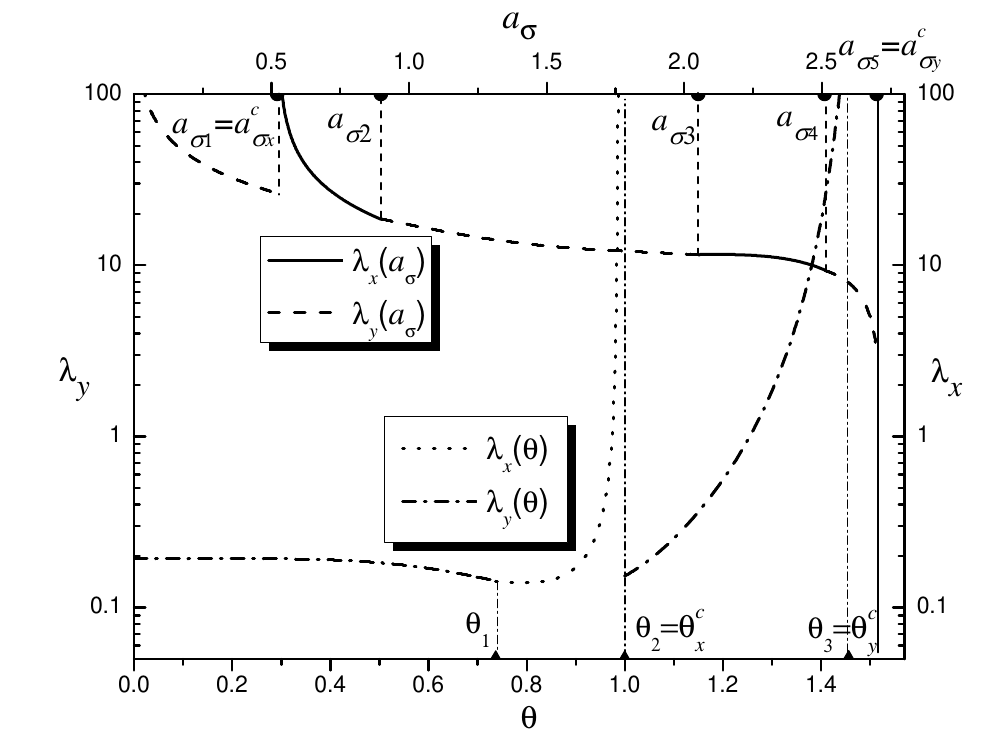}  b) \includegraphics[width=80mm]{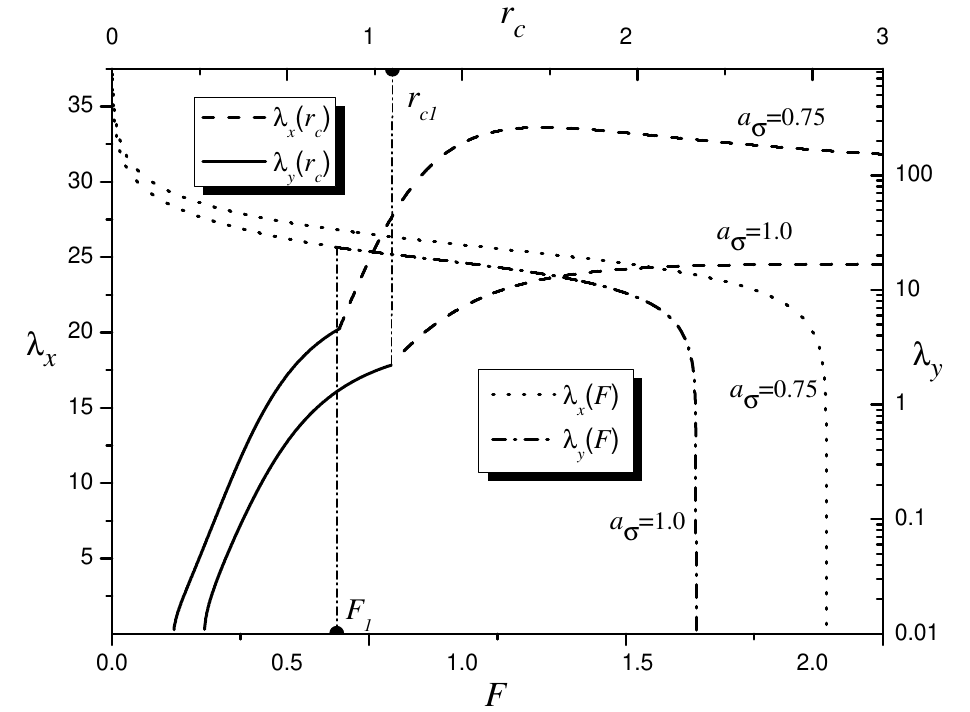}
\caption{ (a) Plot of selected wave-lengths $\lambda_x$ and $\lambda_y$ vs.
incidence angle $\theta$ at $a_\sigma=0.45$, $r_c=0.65$ and dimensionless
penetration depth $a_\sigma$ at $\theta=0.4763$, $r_c=1.0$ ($a_{\sigma i}$ with
$i\in 1,\ldots,4$ denotes threshold values when a change of the wave-vector of
patterns occurs); here $0<\theta<\pi/2$ is measured in radians; other
parameters are: $F=1$, $\sigma=1$. (b) Plot of selected wave-lengths
$\lambda_x$ and $\lambda_y$ vs. the correlation scale $r_c$ at $F=1.0$ and the
energetic parameter $F$ at $r_c=1.0$ at $\theta=0.4763$. Other parameters are:
$F=1$, $\sigma=1$, $\Sigma=1$.\label{fig2}}
\end{figure}

Next, we calculate the selected wave-lengths $\lambda_x$ and $\lambda_y$ versus
the angle of incidence $\theta$ and the penetration depth $a_\sigma$
(Fig.\ref{fig2}a) and versus the correlation scale $r_c$ and the energetic
parameter $F$ (Fig.\ref{fig2}b). The selected wave-lengths relate to the
smallest wave-number in the corresponding direction. It is seen that as
$a_\sigma$ or $\theta$ varies transformations in ripple orientation occur. Here
$a_{\sigma i}$ and $\theta_i$ are threshold magnitudes for the penetration
depth and incidence angle, respectively, indicating change of the ripple
orientation. It is seen that there are two critical values $a_{\sigma
1}=a^c_{\sigma x}$ and $a_{\sigma 5}=a^c_{\sigma y}$ where the corresponding
wave-lengths take infinitely large magnitudes due to $\Gamma_x=0$. There are
two critical value for the angle $\theta_2=\theta_x^c$ and
$\theta_3=\theta_y^c$ indicating divergence of the wave-lengths when $\Gamma_x$
takes zero values. From Fig.\ref{fig2}b one can see that as the energetic
parameter $F$ increases the wave-length of the ripple formation reduces to
zero. At small $a_\sigma$ orientation of selected ripples can be changed at
$F=F_1$, whereas at large values for the penetration depth no change is
possible in the ripple orientation. The dependencies $\lambda_\alpha(r_c)$
manifest non-monotonic behavior: at small $r_c$ the wave-length increases,
whereas at large $r_c$ the decreasing dependencies are observed. Moreover,
there is a critical value for the correlation radius $r_{c1}$ where orientation
of ripples can be changed. Therefore, correlation properties of the ion beam
can play a crucial role in ripple formation processes at early stages (in
linear models). From the equations obtained for the selected wave-numbers it
follows that the selected wave-lengths have the well-known assymptotics versus
main parameters of the beam ($\lambda\sim T^{-1/2}\exp(-E_a/T)$, $\lambda\sim
\epsilon^{-1/2}$, $\lambda\sim J^{-1/2}$) and depend assymptoticaly versus
secondary characteristics: $\lambda\sim (\Sigma_0-\Sigma)^{1/2}$, $\lambda\sim
(r_{c0}-r_c)^{-1}$.

\section{Nonlinear stochastic model}\label{NLIN}

Next, let us consider the nonlinear system behavior setting $\lambda_\alpha\ne
0$. In further study we are based on the simulation procedure allowing us to
solve the nonlinear stochastic differential equation (\ref{eq2}). As it was
done in previous section we use the finite-difference approach to calculate the
evolution of the field $h$.

\subsection{Evolution of the height distribution function}
To investigate properties of a distribution of the field $h$ we use skewness
$m_3$ and kurtosis $m_4$, defined as
\begin{equation}\begin{split}
 &m_3=\frac{\langle (h(\mathbf{r})-\langle{h}(\mathbf{r})\rangle)^3\rangle}{W^3},\\
 &m_4=\frac{\langle
(h(\mathbf{r})-\langle{h}(\mathbf{r})\rangle)^4\rangle}{W^4},\\
&W^2=\langle(h(\mathbf{r})-\langle{h}(\mathbf{r})\rangle)^2\rangle,
\end{split}
\end{equation}  where
$\langle{h}(\mathbf{r})\rangle$ is the average of the height field
($\langle{h}(\mathbf{r})\rangle\equiv V^{-1}\sum_{\mathbf{r}}h(\mathbf{r},t)$,
$V=L^d$ is the system volume, $d$ is the spatial dimension, $L$ is the linear
size of the system); $W$ is the interface width. Skewness is a measure of the
symmetry of a profile about the reference surface level. Its sign tells whether
the father points are proportionately above ($m_3>0$) or below ($m_3<0$) the
average surface level. Kurtosis describes randomness of the surface relative to
that of a perfectly random (Gaussian) surface, for the Gaussian distribution
one has $m_4=3.0$. Kurtosis is a measure of the sharpness of the height
distribution function. It is known that if most of the surface features are
concentrated near the mean surface level, then the kurtosis will be less than
if the height distribution contained a larger portion of the surface features
lying farther from the mean surface level.

\begin{figure}[!t]
\centering
 a) \includegraphics[width=120mm]{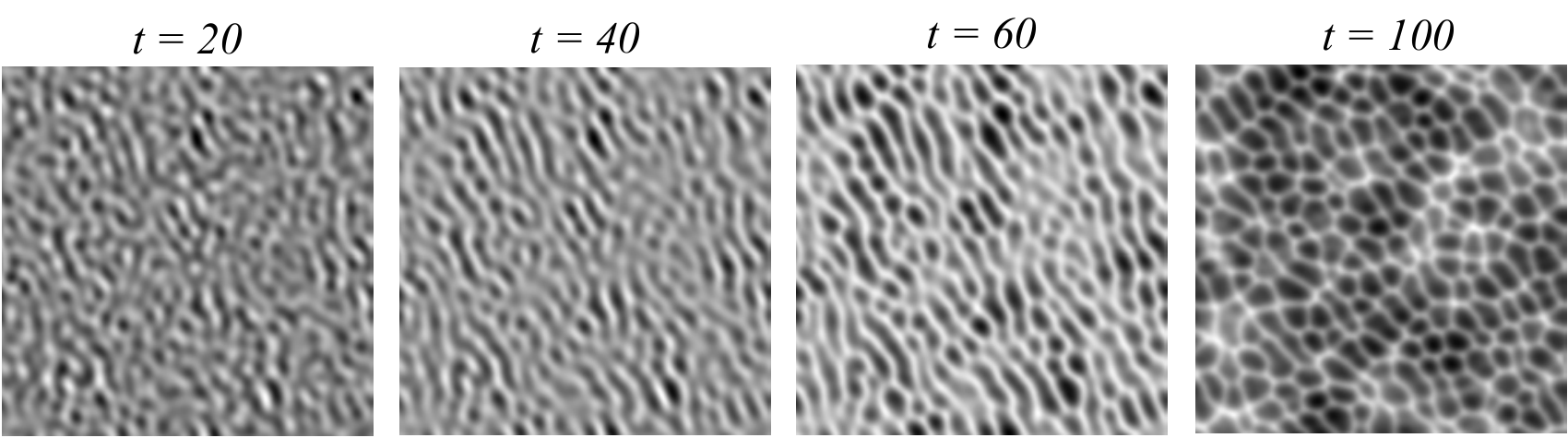}\\
 b) \includegraphics[width=80mm]{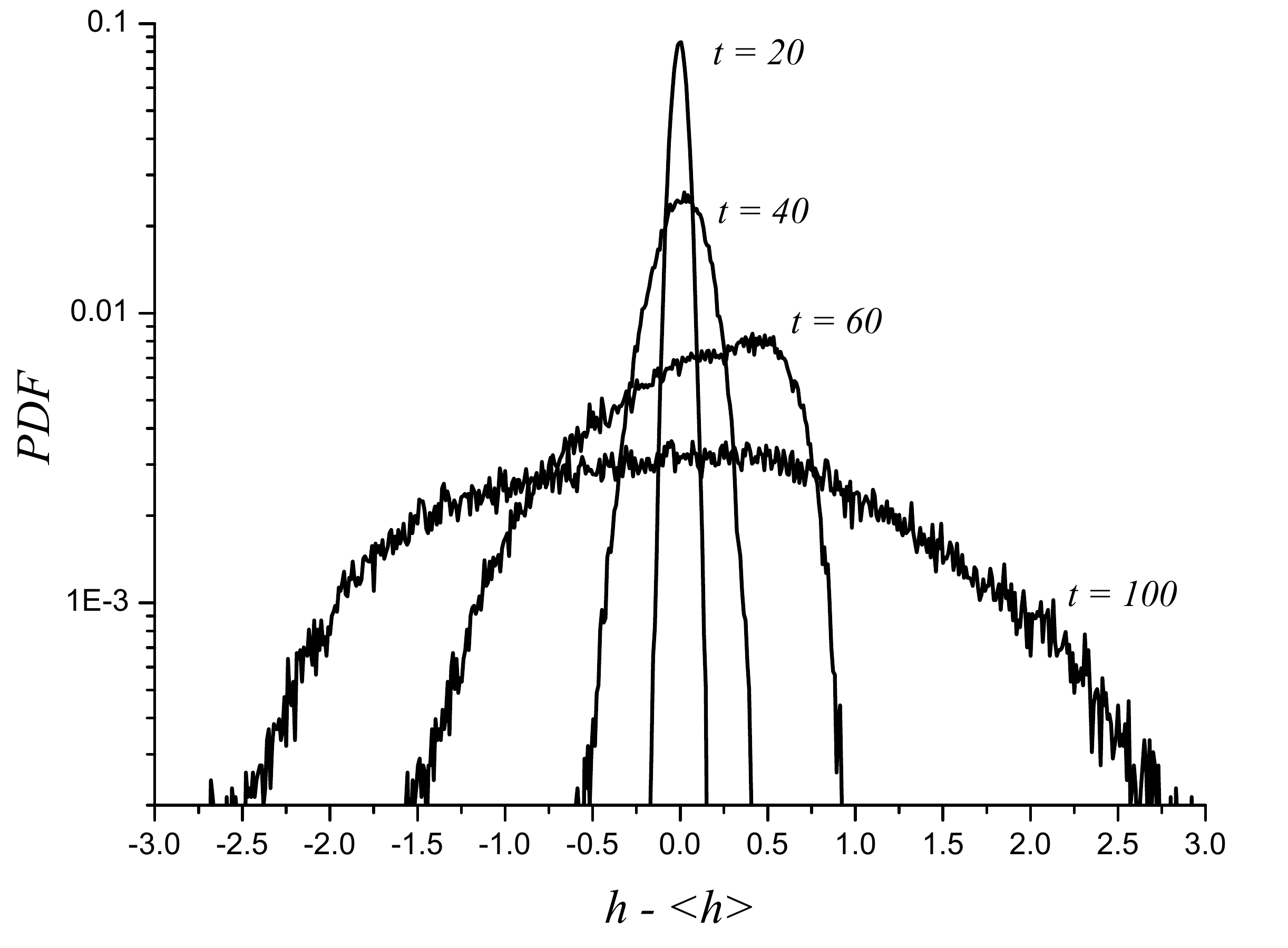}\  c) \includegraphics[width=80mm]{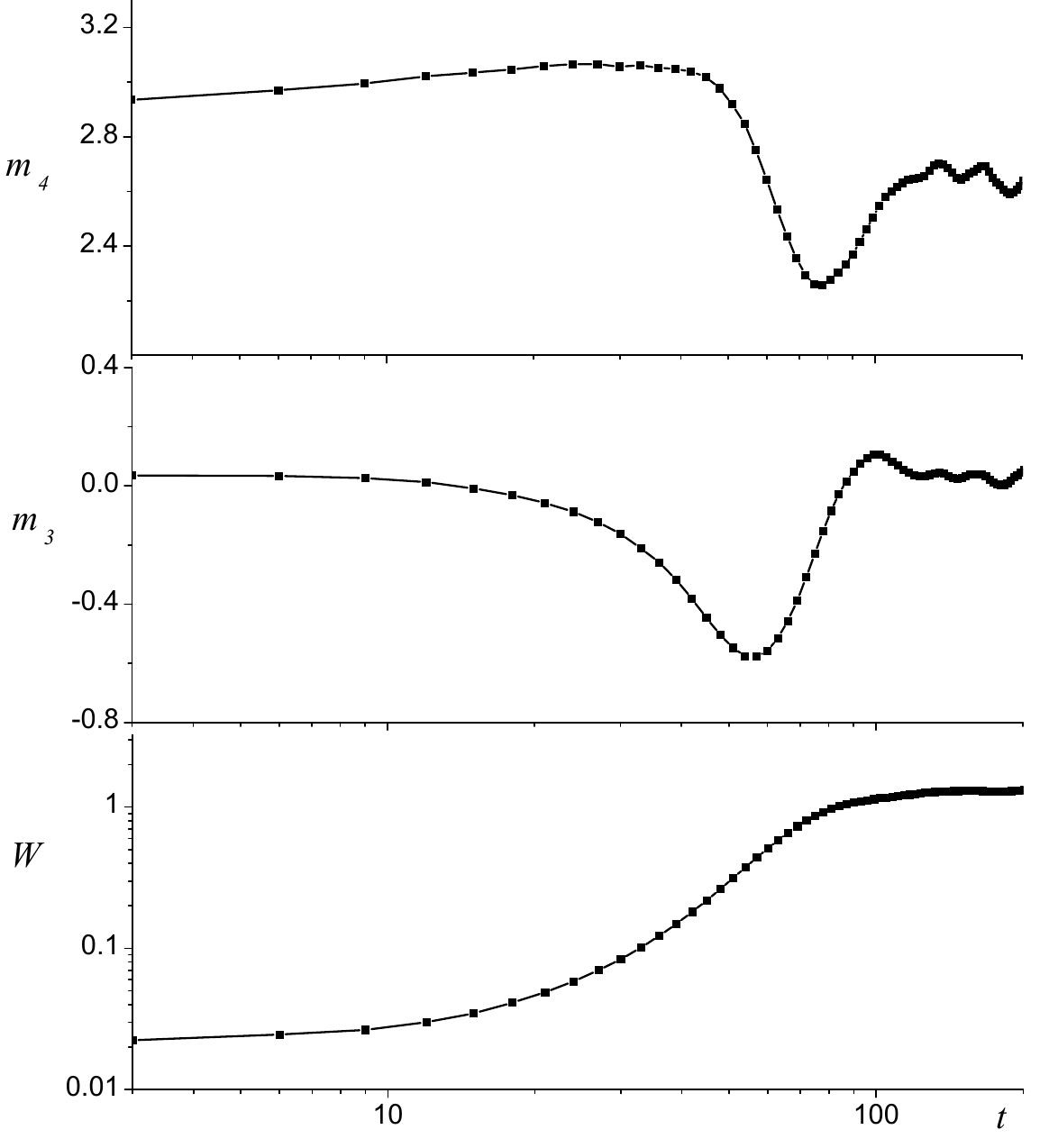}
\caption{A typical evolution of the system with multiplicative fluctuations at
early stages. (a) Snapshots of images of the field $h$ distribution for various
growth times (dark color indicate low $h$, white areas relate to high $h$). b)
Probability density function of the height for various growth time. c)
Kurtosis, skewness, and interface width versus growth time. Other parameters
are: $a_\sigma=1.2$, $\theta=0.4$, $F=1.0$, $\sigma=1.0$, $K=2.0$,
$\Sigma=1.0$, $r_c=1.0$. \label{evol}}
\end{figure}

Figure \ref{evol}a shows snapshots of the surface morphology for the set of
parameters: $a_\sigma=1.2$, $\theta=0.4$, $F=1.0$, $\sigma=1.0$, $K=2.0$,
$\Sigma=1.0$, $r_c=1.0$ at $t=20$, 40, 60, and 100, respectively. In our
simulations we have used Gaussian initial conditions taking $\langle
h(\mathbf{r},t=0)\rangle=0$, $\langle(\delta h)^2\rangle=0.1$; integration time
step is $\Delta t=0.005$, space step is $\ell =1$. It is seen that with an
increase of the growth time, the lateral length of the surface features becomes
bigger and holes (black regions) are formed at $t=100$. It follows that due to
nonlinear effects and noise action the surface morphology is crucially changed
comparing to the ripples shown in Fig.\ref{fig1}b. Figure \ref{evol}b
illustrates the corresponding height probability density distribution function
of these surfaces. It is seen that at $t=20$ the distribution is close to the
Gaussian distribution. With the increase of the growth time, there is deviation
from zero-centered Gaussian distribution and after transient period of time the
probability density function becomes symmetrical and centered around zero. In
Fig.\ref{evol}c we plot the kurtosis $m_4$, the skewness $m_3$, and the
interface width $W$ as functions of the growth time for above system
parameters. According to initial conditions we have $m_4\simeq 3.0$, $m_3\simeq
0$ and $W\simeq 0$ at $t\simeq 0$. With the increase of the growth time the
kurtosis grows until maximum is reached. The skewness decreases to its minimum,
and after tends to zero. These two quantities reflect the form of the
distribution function shown in Fig.\ref{evol}b. The interface width increases
algebraically toward a saturation regime at large $t$.

\begin{figure}
\centering
 \includegraphics[width=100mm]{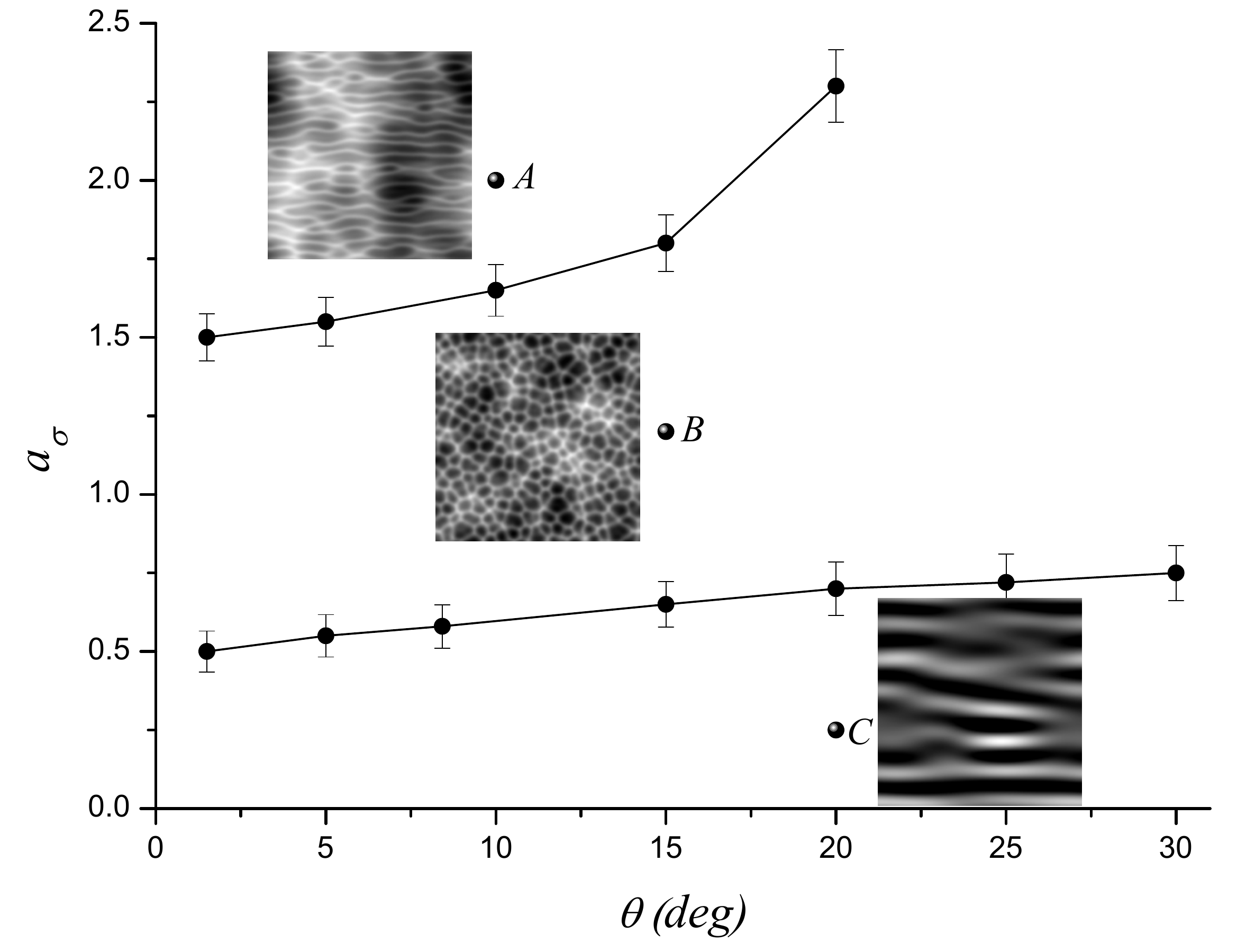}
\caption{Phase diagram for the anisotropic nonlinear model at $F=1$,
$\sigma=1$, $\Sigma=1$, $r_c=1$. Snapshots are taken at the system parameters
related to position of the points $A$, $B$ and $C$,
respectively.\label{phd_nlin}}
\end{figure}

We have computed phase diagram for the nonlinear systems illustrating formation
of different patterns shown in Fig.\ref{phd_nlin}.
It is seen that the numerical results are well related to analytical
predictions from the linear stability analysis. Indeed, critical points lying
on the lines correspond to a change of the sing of the quantity $\Gamma_x$. At
large and low penetration depth $a_\sigma$ ripples oriented along $k_y$
direction are observed (see snapshot for the point $C$) due to $\Gamma_x>0$. At
the intermediate values of $a_\sigma$ random patterns (holes) are realized due
to the nonlinear influence of both the deterministic term
$\lambda_\alpha(\nabla_\alpha h)^2$ and the stochastic contribution.

\subsection{Scaling properties of the surface morphology}

Using numerical data it is possible to study statistical properties of the
system considering the time-dependent height-height correlation function,
determined as follows
$C_h(\mathbf{r},t)=\langle(h(\mathbf{r}+\mathbf{r}',t)-{h}(\mathbf{r}',t))^2\rangle$.
In the framework of dynamic scaling hypothesis one can write the correlation
function in the form \cite{FV85,F90}
\begin{equation}
C_h(\mathbf{r},t)=2W^2(t)\phi\left(\frac{r}{\xi(t)}\right)
\end{equation}
where
\begin{equation}
\phi(u)\sim
   \begin{cases} u^{2\alpha}, &\text{for}  \ \ u\ll 1,\\
	\text{const}, &\text{for}  \ \ u\gg 1.
 \end{cases}
\end{equation}
 Early
stages can be fitted by the function \cite{SSG88} $C_h(\mathbf{r},t)\approx
2W^2(t)[1-\exp[-(r/\xi)^{2\alpha}]$. The dynamic scaling hypothesis assumes
that the following dependencies are hold: $W^2(t)\propto t^{2\beta}$,
$\xi(t)\propto t^{1/z}$, where $\beta$ is the growth exponent, $z$ is the
dynamic exponent for which $z=\alpha/\beta$. From another viewpoint one can
assume \cite{GGR2002}
\begin{equation}\label{C2}
C_h(\mathbf{r},t)=r^{2\alpha}\psi\left(\frac{t}{r^z}\right)
\end{equation}
where
\begin{equation}
\psi(v)\sim
   \begin{cases} v^{2\beta}, &\text{for}  \ \ v\ll 1,\\
	\text{const}, &\text{for}  \ \ v\gg 1,
 \end{cases}
\end{equation}
and the relation $z=\alpha/\beta$ holds. Therefore, these two cases lead to the
same results $C_h(t)\propto t^{2\beta}$ and $C_h(r)\propto r^{2\alpha}$,
allowing one to define the growth exponent $\beta$ and the roughness exponent
$\alpha$. As was shown in Ref.\cite{GGR2002} the roughness $W(t,L)$ can be
related to the structure function $S(\mathbf{k})$ as follows
$W^2(t,L)=V^{-1}\sum_{\mathbf{k}\ne 0 }S(\mathbf{k},t)$, where
$S_h(k,t)=V^{-1}\langle h_k(t)h_{-k}(t)\rangle$. The structure function
$S(k,t)$ has the form
\begin{equation}
S_h(k,t)=k^{-(d+2\alpha)}\Theta(k^zt),
\end{equation}
where
\begin{equation}
\Theta(k^zt)\sim
   \begin{cases} k^{2\alpha} t^{2\alpha/\beta}, &\text{for}  \ \ k^zt\ll 1,\\
	\text{const}, &\text{for}  \ \ k^zt\gg 1,
 \end{cases}
\end{equation}
and scales as $S_h(k,t)\propto k^{-(d+2\alpha)}$ for large $t$ and
$S_h(k,t)\propto t^{2\beta}$ for small $t$.

In previous studies (see, for example Ref.\cite{DZLW99}) it was shown that even
in the isotropic system with additive noise scaling exponents $\alpha$, $\beta$
and $z$ depend on the system parameters $\nu_0$, $\lambda_0$ and $K$. Moreover,
these exponents are the time-dependent functions, i.e. its magnitudes can be
changed in the course of the system evolution.

In our study we have taken into account multiplicative noise described by the
energetic characteristics of the beam and additionally by the noise intensity
$\Sigma$ and correlation radius of fluctuations $r_c$. Therefore, one should
await that due to renormalization of the main system parameters responsible for
the stability of the system and nonlinear effects in its behavior such scaling
exponents are functions of the above noise properties. To prove it we compare
magnitudes of both scaling exponents $\alpha$ and $\beta$ for the system with
additive fluctuations and for the system with our multiplicative noise.

According to the scaling hypothesis the temporal evolution of the quantity
$W=\langle(\delta h)^2\rangle$, where $\delta h=h-\langle h\rangle$, can be
represented through the exponent $\delta$ related to the exponent $\beta$ as
$\delta=\beta$. It is known that the ordinary diffusion (Brownian) process is
described by Einshtein law $\langle(\delta h)^2\rangle\propto t^{2\delta}$,
with $\delta=1/2$. If the exponent $\delta$ deviates from the value 1/2, then
anomalous processes are realized: at $0<\delta<1/2$ there is a delayed
(subdiffusion) process, whereas at $\delta>1/2$ the accelerated diffusion
(superdiffusion) is realized. By comparison of results related to additive and
multiplicative noise influence in anisotropic system, one can see that in the
case of the additive noise influence we get $1/2<\delta<1$. In the case of the
multiplicative noise influence the quantity $\delta$ takes values in the window
$0<\delta<4$. It means that at small time intervals there are delayed processes
which can be accelerated by the noise action at intermediate $t$; at large $t$
a transition toward saturation regime reduces growth velocity, decreasing
$\delta$.

To characterize fractal properties of the surface one can study a pair
correlation function defined as follows:
\begin{equation}\label{C_int}
C_{p}(\mathbf{r};t)=\langle h(\mathbf{r}+\mathbf{r}';t)h(\mathbf{r};t)\rangle.
\end{equation}
If there is no characteristic space scale, then the introduced correlation
function should behave itself algebraically, i.e., $C_{p}(r;t)\propto
1/r^{\Delta}$, where the scaling exponent $\Delta$ relates to the fractal
correlation dimension $D_2$ as $\Delta=d-D_2$. The corresponding Fourier
transformation of the correlation function $C_{p}(\mathbf{r};t)$ scales as
$S_p(k;t)\propto k^{-D_2}$. From the definition of the correlation fractal
dimension $D_2$ and the properties of the Fourier component of the correlator
(\ref{C_int}) it follows that at $D_2=0$ there is no scaling behavior of the
structure function and $S_p(k;t)\approx const$. Hence, the surface at the fixed
time $t$ can be considered as a Gaussian surface with no correlation, i.e.
white noise in space with equal contribution of all wave-numbers $k$; the
corresponding spatial correlator (\ref{C_int}) is reduced to the Dirac
delta-function, $C_p(\mathbf{r})\to \delta(\mathbf{r})$. In the case $D_2=2$
one arrives at typical dependence $S_p(k)\propto k^{-2}$ for diffuse spreading
on the structured (let us say, flat) surface. Here the topological dimension
$d$ equals the fractal dimension $D_2$. Therefore, a variation of the fractal
dimension $D_2$ versus the time indicates a change of the fractal morphology of
the surface from pure uncorrelated Gaussian surface toward well structured
surface having fractal dimension $d=D_2=2$.

In order to study scaling properties of the system under consideration we will
compare all our results obtained with results coming from an investigation of
the anisotropic system with additive fluctuations. Such system will serve as a
reference system. Moreover, to verify and to test our the numeric procedure of
the scaling exponents computation we recalculate results of the work
\cite{DZLW99} for the isotropic Kuramoto-Sivashinsky equation.

\begin{figure}[!t]
\centering
\includegraphics[width=120mm]{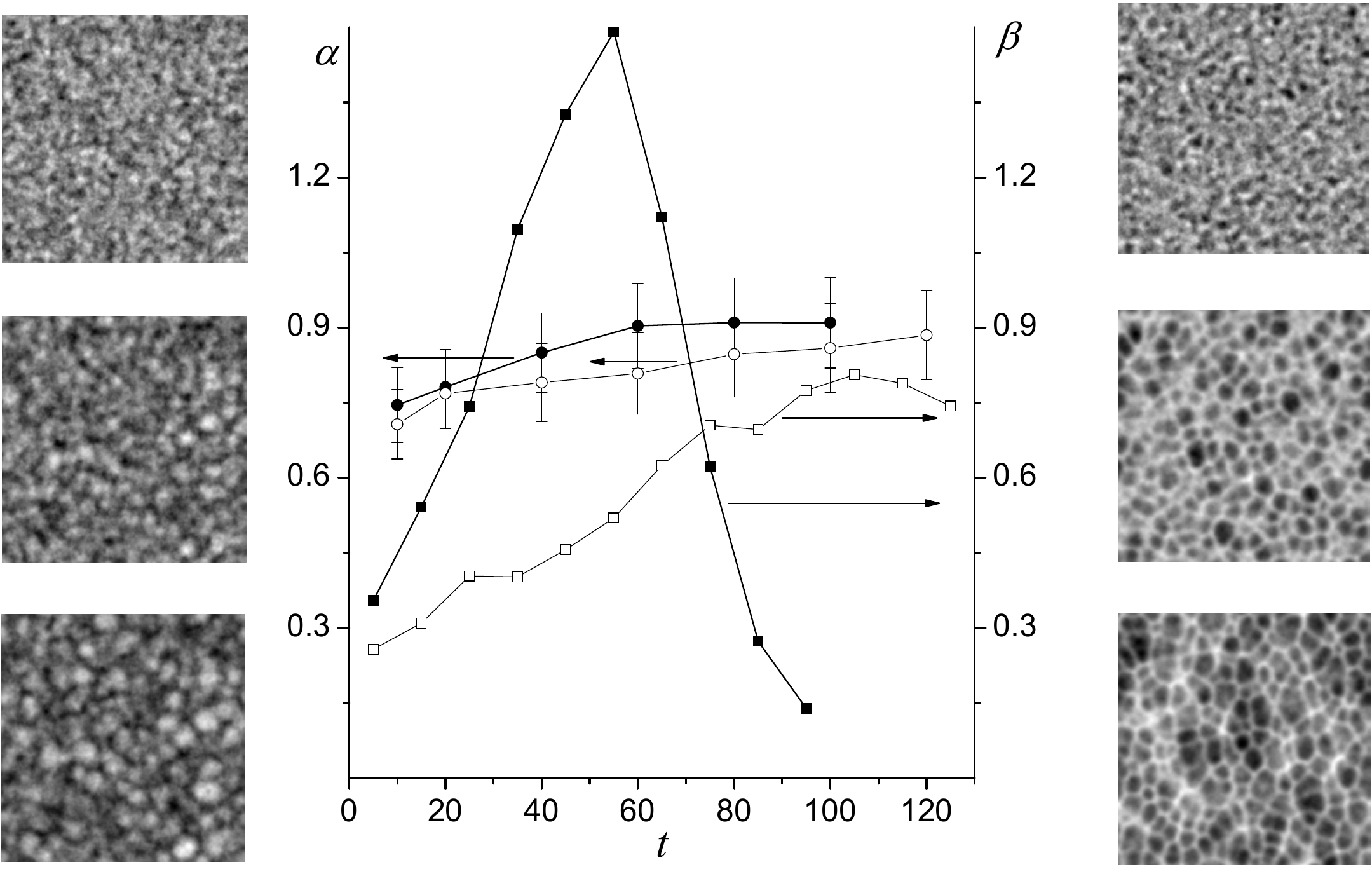}
\caption{Roughness exponent $\alpha$ and growth exponent $\beta$ versus growth
time for isotropic Kuramoto-Sivashinsky equation with additive noise (white
circles and squares) at $\nu_x=\nu_y=-0.2$, $\lambda_x=\lambda_y=1.0$, $K=2$
and anisotropic Kuramoto-Sivashinsky equation with additive noise (black
circles and squares) at $a_\sigma=1.2$, $\theta=0.4$ $K=2$. Snapshots are shown
for above two models (left for isotropic and right for anisotropic
Kuramoto-Sivashinsky equation with additive noise) at $t=20$, 60, 100 from top
to bottom. The noise intensity $\Sigma=1.0$.\label{ad_noise}}
\end{figure}

As a reference system we consider the model described by the Langevin equation
with additive noise, i.e., $\partial_t h=\nu_{\alpha
0}\nabla_{\alpha\alpha}^2h+\frac{\lambda_{\alpha 0}}{2}(\nabla_\alpha h)^2-
K\nabla^2(\nabla^2h) +\zeta(\mathbf{r},t)$, where $\zeta$ is the Gaussian
random source with properties $\langle\zeta\rangle=0$,
$\langle\zeta(\mathbf{r},t)\zeta(\mathbf{r}',t')\rangle=2\Sigma\delta(\mathbf{r}-\mathbf{r}')\delta(t-t')$.
Calculations of the dynamical exponents at the system parameters
$a_\sigma=1.2$, $\theta=0.4$, $F=1.0$, $\sigma=1.0$, $K=2.0$, $\Sigma=1.0$,
$r_c=1.0$ are shown in Fig.\ref{ad_noise}. It is seen that the exponents
$\alpha$ and $\beta$ for above two models are different. In the anisotropic
case we have elevated magnitudes for $\alpha$ and $\beta$, i.e. such exponents
essentially depend on the control parameters of the system. Hence, due to
renormalization of the control parameters by the multiplicative noise
contribution the dynamic scaling exponents depend on the noise characteristics.

Let us consider the anisotropic system with multiplicative fluctuations. We
have performed calculations of the scaling exponents at the fixed point on the
phase diagram ($\theta$,$a_\sigma$) at different values of the noise intensity
$\Sigma$ and the correlation radius $r_c$. The reference point is
$a_\sigma=1.2$, $\theta=0.4$, $F=1.0$, $\sigma=1.0$, $K=2.0$. We compute
$\alpha$ and $\beta$ at time window when the interface width $W$ or the
correlation function $C_h(r)$ start to grow until they saturate (i.e., when
algebraic dependencies $W^2(t)\propto t^{2\beta}$ and $C_h(r)\propto
r^{2\alpha}$ are observed).

\begin{figure}
\centering
 a)\includegraphics[width=160mm]{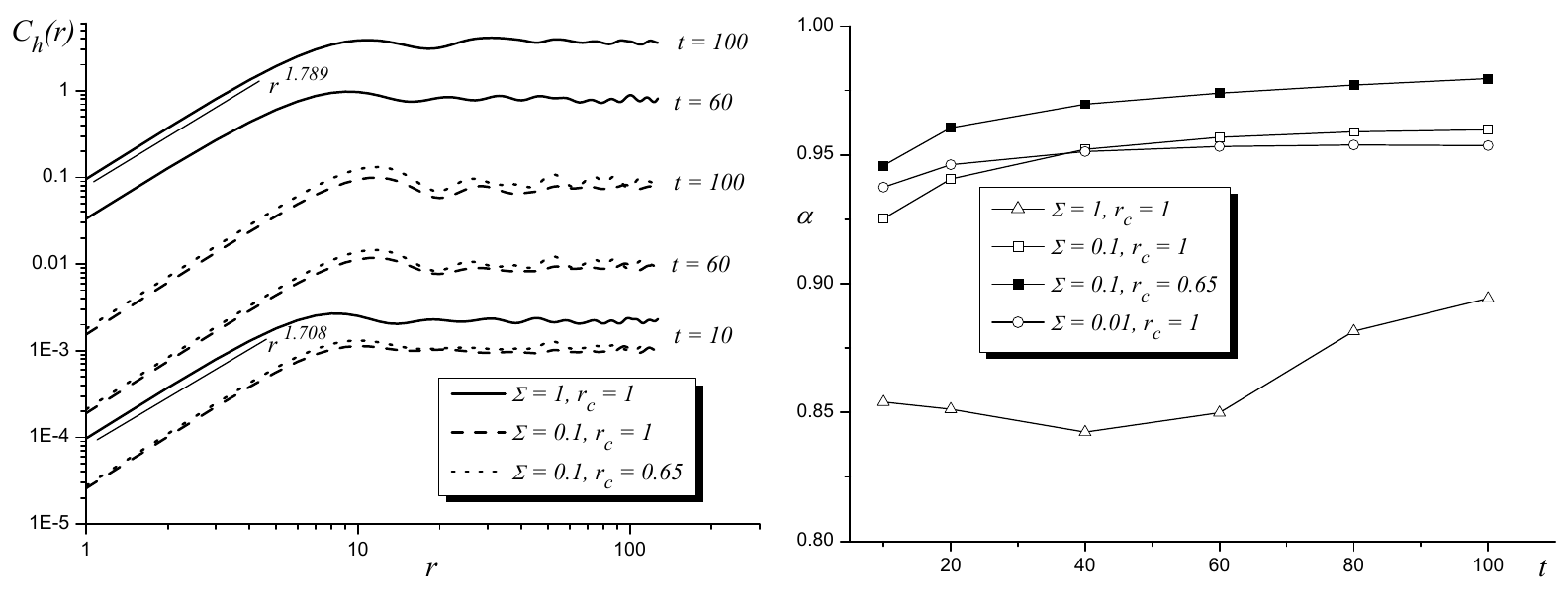}\\
 b)\includegraphics[width=80mm]{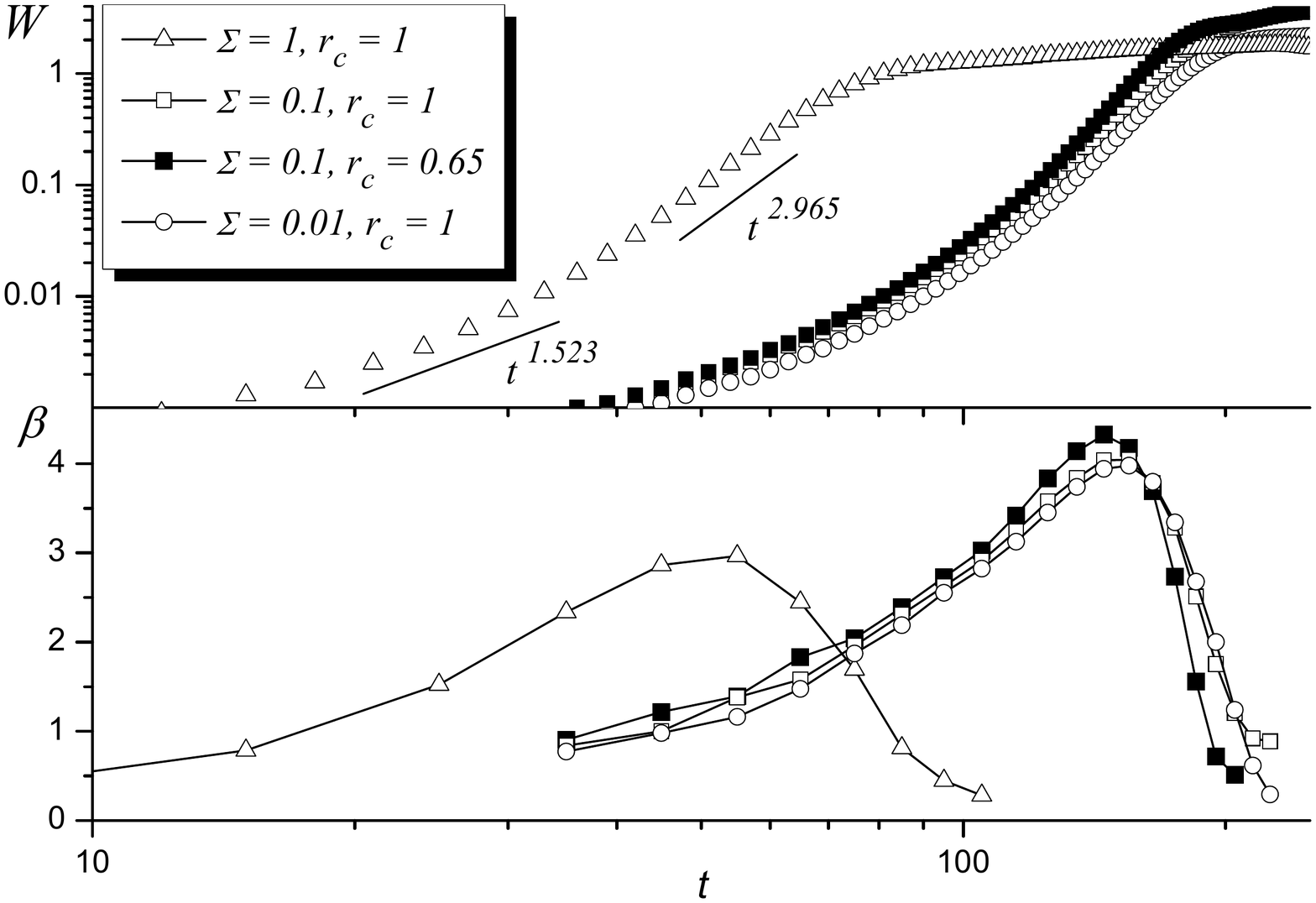}\\
 c)\includegraphics[width=160mm]{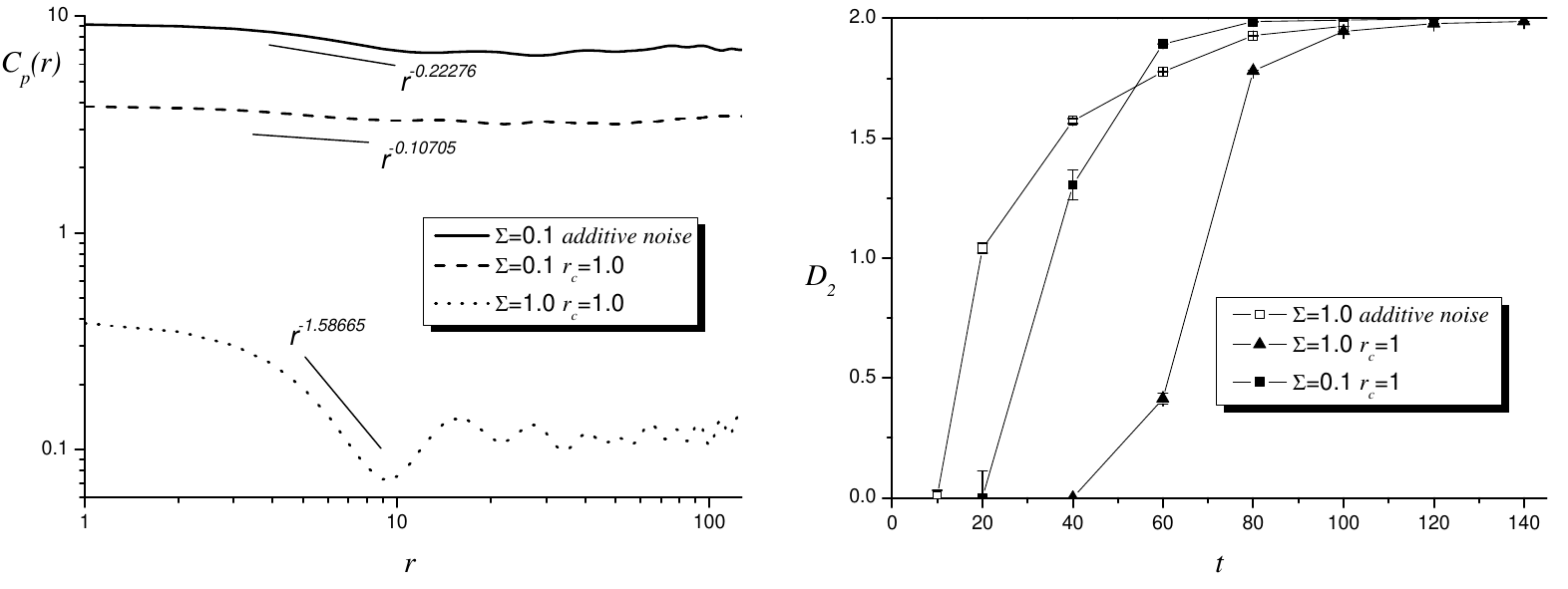}
\caption{(a) The correlation function $C_h(r)$ and the roughness exponent
$\alpha$ versus time. (b) The interface width $W$ and the growth exponent
$\beta$ versus time. (c) The correlation function $C_p(r)$ at $t=60$ and the
fractal dimension $D_2$ versus time. Other parameters are: $a_\sigma=1.2$,
$\theta=0.4$, $F=1$, $\sigma=1$, $K=2$. \label{ab_mult}}
\end{figure}

The corresponding time dependencies of $\alpha$,
 $\beta$ and $D_2$ are shown in Fig.\ref{ab_mult}. In Fig.\ref{ab_mult}a we plot
the corresponding correlation function $C_h(r;t)$ and the roughness exponent
$\alpha$; Figure \ref{ab_mult}b illustrates the time dependence of the
interface width $W$ and the growth exponent $\beta$; Figure \ref{ab_mult}c
shows the pair correlation function $C_p(r;t)$ and the associated fractal
dimension $D_2$ at fixed times.
From Fig.\ref{ab_mult}a it is seen that the growth process is nonstationary for
early stages and the roughness exponent $\alpha$ is near 0.95 for small
incidence angle dispersion $\Sigma$. At such set of the control parameters
($a_\sigma$ and $\theta$) the correlation radius $r_c$ has not essential
influences on the system behavior. At large $\Sigma$ the roughness exponent has
lower magnitudes and $\alpha$ has the well pronounced time dependence.

Comparing curves for the interface width at different $\Sigma$ and $r_c$ from
the one hand and the growth exponents dependencies versus time from another one
(see Fig.\ref{ab_mult}b), one can conclude that as the noise intensity $\Sigma$
increases the position of the peak of the exponent $\beta$ reduces to small
time. It means that as the noise intensity increases at such choice of the
control parameters the interface width increases at smaller time interval than
at low $\Sigma$. Alternatively, the shift of the peak position at large
$\Sigma$ indicates that multiplicative fluctuations are responsible for
nonlinear effects at small times. It looks natural due to the nonlinear form of
the multiplicative noise, where the large noise contribution influences
crucially on properties of the growth processes. The correlation properties of
fluctuations characterized by $r_c$ define a height of the peak for $\beta$. In
other words, the noise correlations can accelerate growth processes increasing
the interface width $W$ until it attains the saturation.

A change of the fractal properties of the surface is shown in
Fig.\ref{ab_mult}c. Here we compare fractal properties of two different
systems, namely with additive fluctuation source and with multiplicative noise.
From the dependencies of the pair correlation function $C_p(r)$ it is seen that
at $t=60$ the additive noise contribution leads to a picture when the
correlation function $C_p(r)$ decreases slowly with exponent $\Delta=0.227$,
whereas the multiplicative noise contribution with the same intensity
$\Sigma=1.0$ at $r_c=1.0$ increases the exponent $\Delta$ to 1.587. According
to the definition of the correlation dimension $D_2$ it means that the fractal
properties of the surface is well pronounced at multiplicative noise with large
intensity at a small time interval $t\simeq 60$ (see curves $D_2(t)$). From the
time dependencies of the fractal dimension $D_2$ for the system with
multiplicative noise it follows that at small times the surface has Gaussian
properties of the kind of white noise in space (the correlation function has
the from of the Dirac delta-function.). At small time interval (at intermediate
times) the fractal properties emerge and characterized by $0<D_2<2$. At large
times one has $D_2=2$ and the well structured patterns are observed, its
dimension $D_2$ coincides with the topological $d=2$. In the case of additive
fluctuations the time interval of the formation of well structured patterns is
larger than in system with multiplicative noise.

\begin{figure}[!t]
\centering a) \includegraphics[width=160mm]{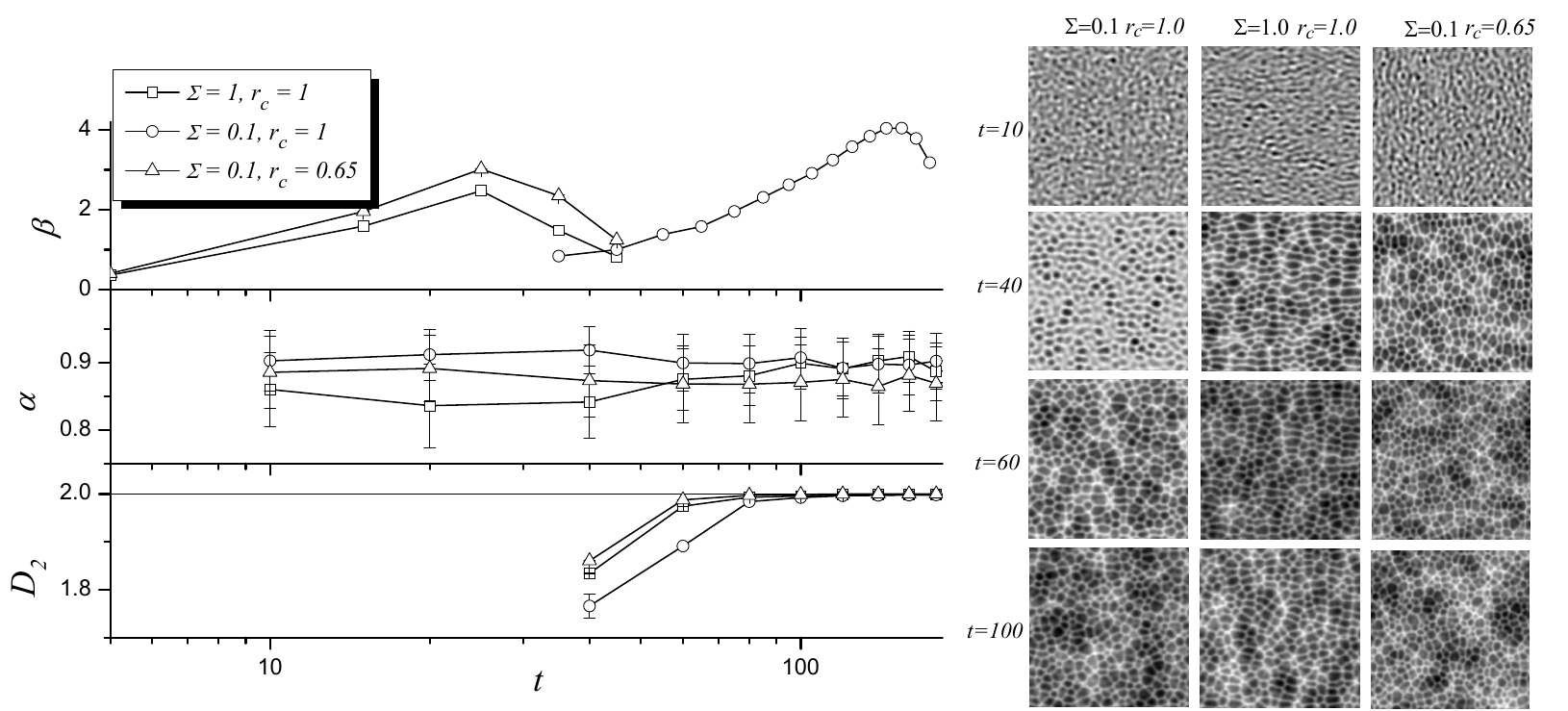}\\
 b)  \includegraphics[width=160mm]{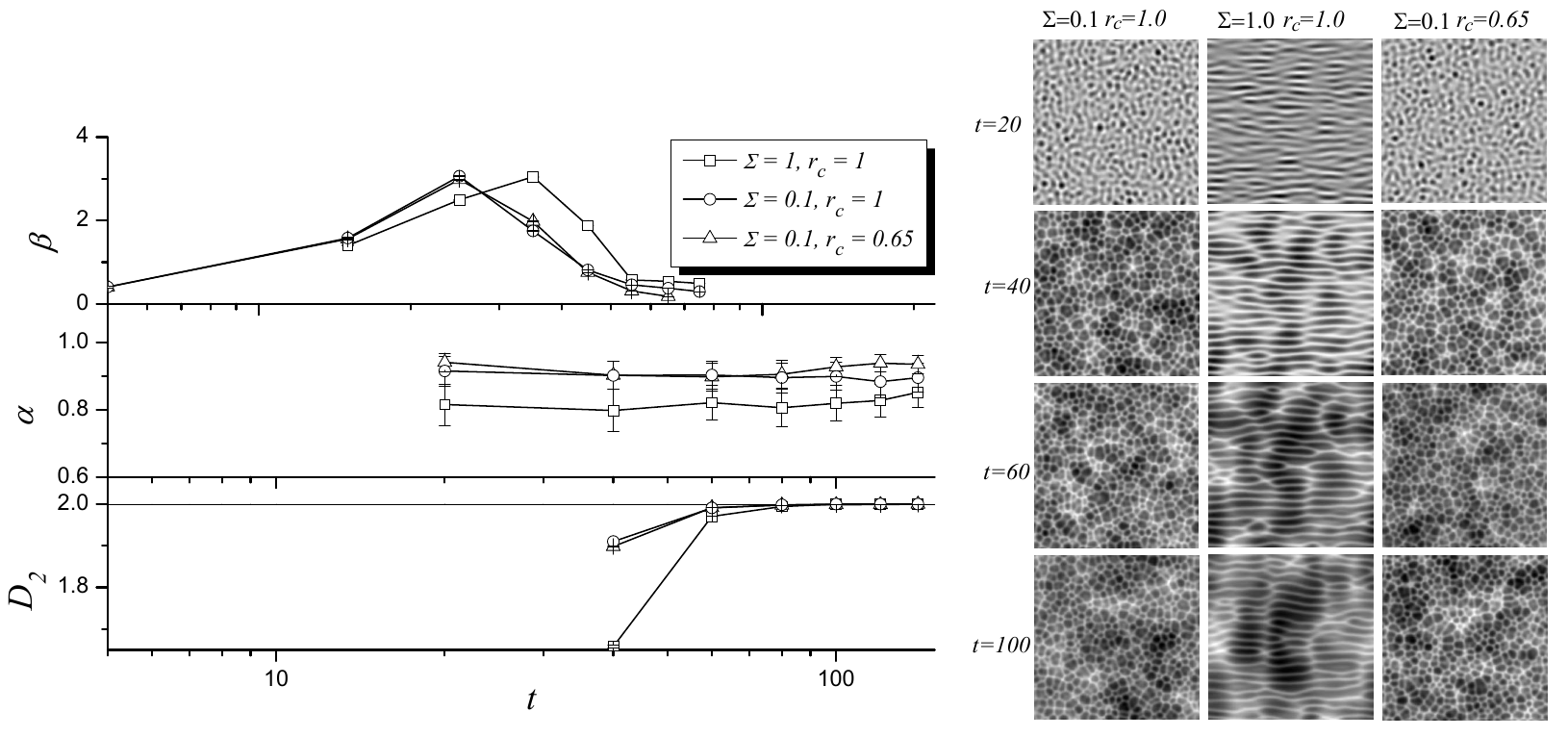}
 \caption{Scaling exponents $\alpha$, $\beta$ and correlation dimension $D_2$
versus time at (a) $a_\sigma=2.0$, $\theta=0.4$ and (b) $a_\sigma=2.0$,
$\theta=0.2$. Typical snapshots of the system evolution are taken at different
$\Sigma$ and $r_c$. Other parameters are: $F=1$, $\sigma=1$,
$K=2$\label{ab_mult2}}
\end{figure}

Next let us compare the time dependencies for the scaling exponents for
different set of the system parameters $a_\sigma$ and $\theta$ shown in
Figs.\ref{ab_mult2}a,b. It is seen that at $a_\sigma=2.0$, $\theta=0.4$
(Fig.\ref{ab_mult2}a) at large noise intensity $\Sigma$ the interface width
grows faster comparing with the case of small $\Sigma$ at fixed $r_c$. But the
growth at small $\Sigma$ occurs at earlier periods of time. This situation
absolutely different to the case shown in Fig.\ref{ab_mult}b, whereas
variations in $r_c$ leads to the increase of the maximum for $\beta$. The
roughness exponent $\alpha$ does not principally change its values at different
$t$. Comparing curves for the fractal dimension $D_2$ for above two sets of the
system parameters one can see that at small $\Sigma$ the quantity $D_2$ is
smaller than in previous case (cf. dependencies $D_2(t)$ in
Figs.\ref{ab_mult}c, \ref{ab_mult2}a). The same dependencies of the scaling
exponent $\alpha$, $\beta$ and $D_2$ at $a_\sigma=2.0$, $\theta=0.2$ are
observed. Here at large $\Sigma$ the nonlinear effects delayed. Considering
snapshots shown in the right hand side of the plots at different $\Sigma$ and
$r_c$ one can see that depending on the noise properties the morphology of the
surface changes crucially (cf. snapshots at different $r_c$ in
Fig.\ref{ab_mult2}b). It means that the phase diagram shown in
Fig.\ref{phd_nlin} will be modified under variation of $\Sigma$ or $r_c$ as
linear stability analysis predicts.

\begin{figure}
\centering
 a)\includegraphics[width=50mm]{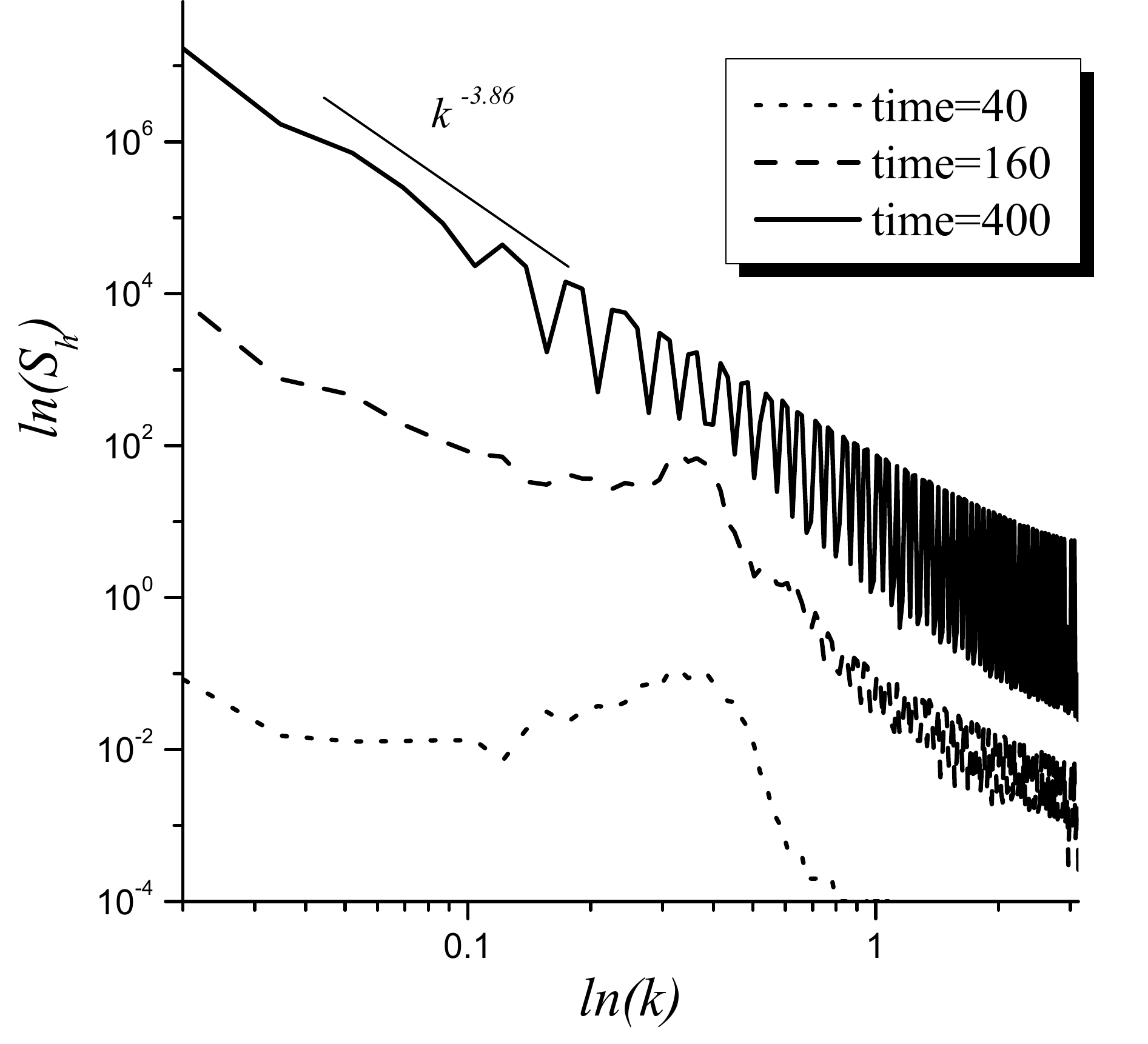}
 b) \includegraphics[width=50mm]{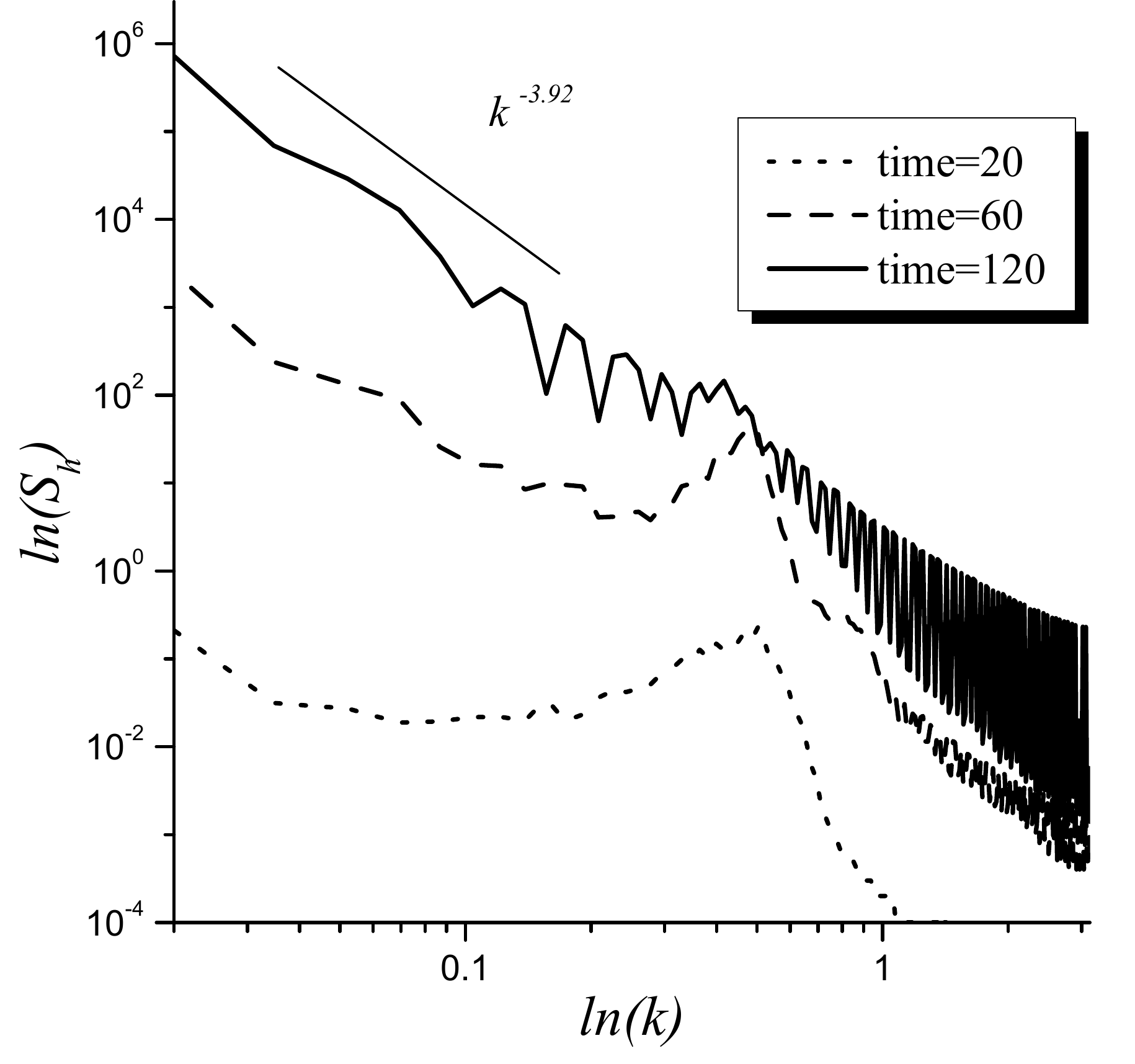}
 c) \includegraphics[width=50mm]{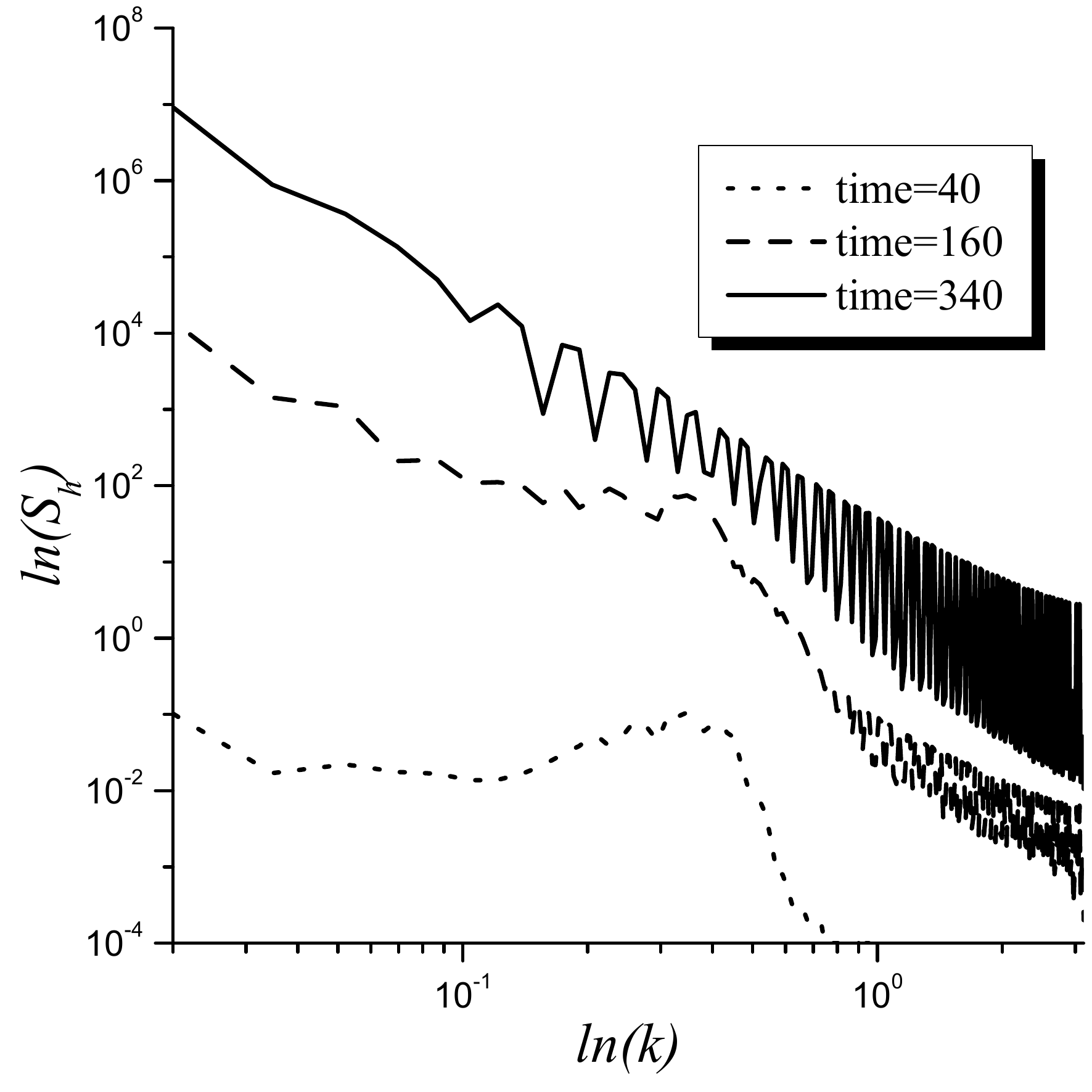}
 \caption{Evolution of the spherically averaged structure function $S_h(k,t)$ at
different noise intensity: a) $\Sigma=0.1$, $r_c=1$; b) $\Sigma=1.0$, $r_c=1$;
c)$\Sigma=0.1$, $r_c=0.65$. Other parameters are: $a_\sigma=1.2$, $\theta=0.4$,
$F=1$, $\sigma=1$.\label{evolS(k)}}
\end{figure}

Figure \ref{evolS(k)} show the evolution of the spherically averaged structure
function defined on a circle
\begin{equation}S_h(k,t)=\frac{1}{N_k}\sum\limits_{k\le
\mathbf{k}\le k+\Delta k}S_h(\mathbf{k},t),
\end{equation}
where $N_k$ is the number of point on the circle of the width $\Delta k$. Two
different choice of the noise intensity $\Sigma$ and the correlation radius of
fluctuations $r_c$ are shown in Figs.\ref{evolS(k)}a,b,c, respectively. It is
seen that during the system evolution at early stages the system selects the
ripples with the corresponding wave-number (dotted lines) and after at late
stages the algebraic form for the structure function is realized. At large time
intervals one can define the exponent $\alpha$ from the definition
$S_h(k)\propto k^{-(d+2\alpha)}$. As is shown in Figs.\ref{evolS(k)}a,b for
above system parameters one has $S_h(k)\propto k^{-3.86}$ for $\Sigma=0.1$ and
$S_h(k)\propto k^{-3.92}$ for $\Sigma=1.0$, where $d=2$. Hence, the roughness
exponent takes values $\alpha\simeq 0.93$ and $\alpha\simeq 0.96$ that is well
predicted by the analysis of the correlation function $C_h(r)$ (see
Figs.\ref{ab_mult}a).

\section{Conclusions}\label{sum}

We have studied the ripple formation processes induced by the ion sputtering
under stochastic conditions of illumination. The main assumption was stochastic
nature of the ion beam when the angle of incidence distributed in space and
time (homogeneous and stationary field). It allows us to generalize the
Bradley-Harper model of ripple formation \cite{BH88} and consider the
stochastic model with multiplicative fluctuations describing random nature of
the incidence angle proposed in Ref.\cite{KYH2009}. We have discussed
properties of the ripple formation in both linear and nonlinear models.

Within the framework of the linear stability analysis we have shown that even
in the linear system the noise action is able to change the critical values for
the control parameters of the system such as the penetration depth and the
averaged incidence angle. It was found that as correlation properties of such
multiplicative noise as the dispersion in the incidence angles around the
average can reduce the domains of the control parameters where the ripples
change their orientation at the fixed angle of incidence.

Studying the nonlinear model we have computed the dynamic phase diagram
illustrating formation of different patterns (ripples and holes) which relates
to the results from the linear stability analysis. Main properties of the
ripple formation were studied with the help of the distribution function of the
height and its averages reduced to the skewness, kurtosis and interface width
(dispersion). To make a detailed analysis of the ripple formation we have
examined scaling behavior of main statistical characteristics of the system
reduced to the correlation functions and its Fourier transforms (structure
functions). It was shown that as the growth and roughness exponents depend on
the control parameters and are time-dependent functions (it was predicted by
previous study of the isotropic Kuromoto-Sivashinsky equation \cite{DZLW99});
these exponents depend on the noise properties: its intensity and the spatial
correlation radius. Comparing results for the system with additive and
multiplicative fluctuations it was shown that multiplicative noise can
crucially accelerate processes of ripple formation, increasing the growth
exponent. As far as our system is anisotropic the noise action is different at
different set of the main control parameter values. Studying fractal properties
of the surface we have calculated the fractal (correlation) dimension as the
time-dependent function. It was shown that in the system with multiplicative
noise the fractal properties appear at small time interval of the surface
growth, whereas in the system with the additive noise this time interval is
wider. These results are well predicted by the correlation functions analysis
and by Fourier transformation of the numerically calculated surface.

Therefore, as patterning as the scaling behavior of the system can be
controlled by additional set of parameters reduced to the dispersion of the
angle of incidence and the correlation properties of its fluctuations.

\end{document}